\documentclass[epjc3]{svjour3}  
\RequirePackage{graphicx}
\usepackage{amsmath}
\usepackage{amssymb}
\RequirePackage[numbers,sort&compress]{natbib}
\journalname{Eur. Phys. J. C}
\begin{document}

\title{Does Chaplygin gas have salvation?}


\author{Juliano P. Campos\thanksref{e1,addr1} \and J\'{u}lio C. Fabris\thanksref{e2,addr2} \and Rafael Perez\thanksref{e3,addr2} \and Oliver F. Piattella\thanksref{e4,addr2} \and Hermano Velten\thanksref{e5,addr3}}

\thankstext{e1}{e-mail: jpcampospt@gmail.com}
\thankstext{e2}{e-mail: fabris@pq.cnpq.br}
\thankstext{e3}{e-mail: rperez@cbpf.br}
\thankstext{e4}{e-mail: oliver.piattella@pq.cnpq.br}
\thankstext{e5}{e-mail: velten@physik.uni-bielefeld.de}


\institute{Centro de Ci\^{e}ncias Exatas e Tecnol\'ogicas - UFRB, Campus Cruz das Almas, 710, 44.380-000, Cruz das Almas, BA, Brazil\label{addr1}
           \and
           Departamento de F\'{i}sica - CCE - UFES, Campus Goiabeiras, 514, 29075-910 Vit\'{o}ria, ES, Brazil\label{addr2}
           \and
           Fakult\"at f\"ur Physik, Universit\"at Bielefeld, Postfach 100131, 33501 Bielefeld, Germany\label{addr3}
}

\date{Received: date / Accepted: date}

\maketitle

\begin{abstract}
We investigate the unification scenario provided by the generalised Chaplygin gas model (a perfect fluid characterized by an equation of state $p = -A/\rho^{\alpha}$). Our concerns lie with a possible tension existing between background kinematic tests and those related to the evolution of small perturbations. We analyse data from the observation of the differential age of the universe, type Ia supernovae, baryon acoustic oscillations and the position of the first peak of the angular spectrum of the cosmic background radiation. We show that these tests favour negative values of the parameter $\alpha$: we find $\alpha = - 0.089^{+0.161}_{-0.128}$ at the 2$\sigma$ level and that $\alpha < 0$ with 85\% confidence. These would correspond to negative values of the square speed of sound which are unacceptable from the point of view of structure formation. We discuss a possible solution to this problem, when the generalised Chaplygin gas is framed in the modified theory of gravity proposed by Rastall. We show that 
a fluid description within this theory does not serve the purpose, but it is necessary to frame the generalised Chaplygin gas in a scalar field theory. Finally, we address the standard general relativistic unification picture provided by the generalised Chaplygin gas in the case $\alpha = 0$: this is usually considered to be undistinguishable from the
standard $\Lambda$CDM model, but we show that the evolution of small perturbations, governed by the M\'esz\'aros equation, is indeed different and the formation of sub-horizon GCG matter halos may be importantly affected in comparison with the $\Lambda$CDM scenario.
\end{abstract}


\section{Introduction}

Under the assumption of homogeneity and isotropy of the universe at large scales, and that General Relativity is the correct theory of gravity, the accelerated expansion \cite{Riess:1998cb, Perlmutter:1998np, komatsu, caldwell} is due to an exotic component, with negative pressure, called dark energy (DE). Moreover, structures such as galaxies and clusters of galaxies seem to require the existence of another exotic component, dubbed dark matter (DM) \cite{bertone, padma}, in order to form. DE and DM constitute the so-called dark sector of the universe, accounting for about $95\%$ of the total cosmic energy budget. Their nature is still a mystery, but they clearly do not fit in the standard model of elementary particles \cite{bertone, padma}.
\par
One important proposal describing the dark sector of the universe is the unification scenario, where DM and DE are considered different manifestations of a unique component. The paradigm of such idea is the Chaplygin gas (GC) model \cite{ugo}, where an equation of state inspired by string theory \cite{jackiw} allows to obtain a dynamic behaviour mimicking DM in the past and DE in current time, as structure formation and the accelerated expansion require. Later, the CG has been phenomenologically generalized, leading to the so-called generalized Chaplygin gas (GCG) \cite{sen, neven}.
\par
Though the idea is very appealing, the GCG faces crucial drawbacks when the model undergoes observational tests. This has been first indicated in reference \cite{ioav}, concerning the behaviour of small perturbations and structure formation, and pointed out again in \cite{finelli, orfeu, Gorini:2007ta, Piattella:2009da, zimdahlnewtonian, zimdahl}. One important aspect is that the so-called kinematic tests (those based on the background expansion and the calculation of distances) seem to favour values for the GCG equation of state parameters that imply negative square speed of sound \cite{colistete, velten} (though they do not exclude positive values). If this is really the case, up to our knowledge in the literature there are only two ways out: $i)$ to introduce {\it ad hoc} entropic perturbations \cite{rrrr, Amendola:2005rk} which nullify (or make positive) the effective speed of sound or $ii)$ to modify the gravity theory, by supposing e.g. a non-conservative theory of gravity \cite{thais}.
\par
Our goal in this work is to understand to which extent the results of \cite{colistete, velten} are really robust from the point of view of the kinematic tests. This question has already been addressed in \cite{velten}, using the differential age of the universe (i.e. the $H(z)$ test) and type Ia Supernova (SNIa). We extend that analysis by enlarging the number of observational data for the $H(z)$ test and by including the position of the first acoustic peak of the cosmic microwave background (CMB) spectrum and the baryonic acoustic oscillation (BAO) tests. We revisit the constraints placed by these datasets, but we do not touch delicate issues such as the calibration problem of the SNIa data \cite{ioavbis}. Our results seem to reveal the robustness of the kinematic tests for the GCG model: we find $\alpha = - 0.087^{+0.159}_{-0.135}$ (at the 2$\sigma$ level). Therefore, we not only confirm the previous results, but stress that negative values of $\alpha$ seem to be favoured over the positive ones (for 
comparison, see Table~1 of \cite{velten}).
\par
Note that we cannot report a real ''tension`` between background tests and the ones involving perturbations, since at 2$\sigma$ we find that positive values of $\alpha$ are still allowed ($\alpha < 0.072$, whereas in \cite{ioav} the authors find $|\alpha| \lesssim 10^{-5}$, for perturbations), but we can speculate how the Chaplygin gas could be ''saved`` if such tension would appear in the future, given the fact that many experiments are ongoing and others forthcoming (such as EUCLID), thus the precision on the observational data is increasing day after day.
\par
Within the ''tension`` scenario, if the GCG is indeed the matter component behind the dark sector, it must get its origin from a different theory of gravity rather than GR. For example, non-standard theories of gravity such as Rastall's \cite{Rastall:1973nw, Batista:2011nu, Fabris:2011wz, Daouda:2012ig, Fabris:2012hw}, $f(R)$ \cite{Sotiriou:2008rp} or Horava-Lifshitz \cite{Horava:2009uw, Ali:2011sv} could possibly improve the status of the GCG. Our discussion in Sec.~V is devoted to open a new window for GCG based cosmological models where standard GR is replaced by Rastall's theory.
\par
We also stress that, in general, the accepted idea is that the GCG (based on GR theory) works only if its parameter space is reduced to the $\Lambda$CDM one, in particular if $\alpha = 0$, see Eq.~(\ref{pressure}) and Eq.~(\ref{Hchap}) \cite{park}. We discuss this issue in Sec.~VI. We point out that the reduction $\alpha = 0$ implies that the GCG does not behave as a pressureless fluid at high redshifts, but as a fluid with a small negative constant pressure. Hence the analysis of the growth of sub-horizon GCG dark matter halos during the matter dominated epoch and the ``M\'esz\'aros effect'' could be different from the standard $\Lambda$CDM picture, possibly leading to a different non-linear clustering pattern at small scales and to a modification of the transfer function. This argument agrees with the conclusions of \cite{julioalpha0}. For a opposite point of view, i.e. that the perturbations in the GCG with $\alpha = 0$ are the same (at all perturbative orders) as in the $\Lambda$CDM, see \cite{martins}.
\par
The paper is organized as follows. In Sec.~\ref{Sec:succandprob}, we review briefly the possible tensions within GCG model tests. In Sec.~\ref{Sec:bgtests} we describe the observational tests we perform, while in Sec.~\ref{Sec:AnandRes} we carry out the statistical analysis based on such tests. In Sec.~\ref{Sec:Healing} we investigate a Rastall's theory approach which could possibly save the GCG. In Sec.~\ref{Sec:meszeq} we discuss the general relativistic limit $\alpha \to 0$ of the GCG and solve its M\'esz\'aros equation. In Sec.~\ref{Sec:Conc} we present our conclusions.

\section{Successes and problems of the Chaplygin gas model}\label{Sec:succandprob}

The GCG model is characterized by the equation of state \cite{ugo, sen, neven}
\begin{equation}\label{pressure}
p_c = - \frac{A}{\rho_c^\alpha}\;,
\end{equation}
where $A$ and $\alpha$ are free parameters. The original Chaplygin gas model, which is somehow connected with the Nambu-Goto action of string theory, implies $\alpha = 1$ \cite{jackiw}, the case
$\alpha \neq 1$ being a phenomenological generalization. Integrating the conservation equation for the fluid,
\begin{equation}
\frac{d\rho_c}{da} + \frac{3}{a}\rho_c(1+w_{c}) = 0\;, \hspace{0.5cm} {\rm with\,\, EoS\,\, parameter}\hspace{0.5cm} w_{c} \equiv \frac{p_c}{\rho_c}=-\frac{A}{\rho_c^{1+\alpha}}\;,
\end{equation}
leads to the following expression for the GCG density as function of the scale factor:
\begin{equation}\label{rhoc}
\rho_c (a)= \rho_{c0}\left[\bar A + \frac{(1 - \bar A)}{a^{3(1 + \alpha)}}\right]^\frac{1}{1 + \alpha}\;,
\end{equation}
where we have defined the new parameter $\bar{A} \equiv A/\rho_0^{\alpha + 1}$ and present-time quantities are indicated by the subscript $0$. From Eq.~(\ref{rhoc}) one can see that the original motivation behind the GCG, i.e. a fluid that evolves from the matter behaviour at early times to a constant density at late times, occurs only if $\alpha\geq - 1$.
\par
When one consider a model where the energy content is given by radiation and baryons, besides the Chaplygin gas, Friedmann's equation becomes
 \begin{eqnarray}\label{Hchap}
\frac{H(z)^2}{H_0^2} &=& \biggr[\Omega_{b0}(1 + z)^3 + \Omega_{r0}(1 + z)^4 + \Omega_{c0}\left(\bar A + (1 - \bar A)(1 + z)^{3(1 + \alpha)}\right)\biggl]^\frac{1}{1 + \alpha}\;,\\
\Omega_{c0} &=& 1 - \Omega_{b0} - \Omega_{r0}\;,
\end{eqnarray}
where $H_0$ is the Hubble constant and $z$ is the redshift, which is related to the scale factor by $a = 1/(1 + z)$. We have also assumed a spatially flat background, according to the recent results of WMAP7 \cite{komatsu}. The Hubble parameter today can be expressed as $H_0 = 100 h$ km s$^{-1}$ Mpc$^{-1}$, where $h \approx 0.7$ (we will adopt the range $0 \leq h < \leq 1$ as a prior for our Bayesian analysis). We also introduce $E(z) \equiv H(z)/H_0$, for future convenience.
\par
The idea of DM-DE unification into a single fluid in expression (\ref{Hchap}), faces difficulties from different points of view. From the theoretical side, if vacuum energy is not responsible for DE, it should be explained how it contributes to gravity and therefore how it enters Eq.~\eqref{Hchap}. From the observational point of view, introducing from Eq.~(\ref{pressure}) the GCG speed of sound
\begin{equation}
 c_{s}^2(a) = -\alpha w_c(a)\;, \hspace{0.2cm}{\rm if\,\,} a=1 \hspace{0.2cm}\rightarrow
\hspace{0.2cm}  c_{s0}^2 = \alpha\frac{A}{\rho_{c0}^{\alpha + 1}} = \alpha\bar A\;,
\end{equation}
one can see that if background tests favour negative values of the parameter $\alpha$, then $c_s^2$ is negative. This is not only an undesirable feature of the model, but depending on how much it is negative a ``tension'' between background and perturbative tests may arise. These problems can be alleviated if, for example, positive entropic perturbations $\sigma = \partial p/\partial S >0$ are introduced such that the effective speed of sound becomes a positive quantity. However, the introduction of entropic perturbations implies new free parameters: there are some degrees of arbitrariness, and in some cases, such as those of reference \cite{rrrr}, it is possible only a slight extension to negative $\alpha$. Another possibility of achieving negative values of $\alpha$ is to implement a scalar version of the GCG model using a non-conservative theory of gravity, like Rastall's theory. This has been done in reference \cite{thais}, and in fact negative values of $\alpha$ seem
still to be favoured at the level of perturbations.
\par
This problem concerns not only the confrontation with the power spectrum data, but also the analysis of the anisotropy of the cosmic microwave background radiation (CMB). In references \cite{finelli, orfeu}, the confrontation of the GCG model with the CMB spectrum has been performed, indicating that the most favoured scenario implies $\alpha \to 0$, that is, the GCG model reduces to the $\Lambda$CDM model. This result is one of the reasons for speculating that the GCG model should be be ruled out.

\section{The background tests}\label{Sec:bgtests}

There are four main background tests for a cosmological model:
\begin{enumerate}
\item The differential age of old galaxies, given by $H(z)$.
\item The SN Ia data.
\item The position of the first CMB acoustic peak.
\item The peak position of the baryonic acoustic oscillations (BAO).
\end{enumerate}
\par
For the differential age data, connected with the evaluation of the age of old galaxies that
have evolved passively, leading to values of $H(z)$ for specific redshifts, there are 13 observational data \cite{verde, stern, verdebis, ma, mabis, Moresco:2012jh}. Recently, a compilation of 21 data points has been considered \cite{ratra}. The fundamental relation is
\begin{equation}
H(z) = - \frac{1}{1 + z}\frac{dz}{dt}\;.
\end{equation}
The value of the Hubble parameter today obtained by the HST ($H_0 = 72$ km s$^{-1}$ Mpc$^{-1}$) could also be added to this sample, but this would imply a prior on the final parameter estimation and, therefore, will not be included here. The analysis with the sample of 13 data points for the Chaplygin gas model has been made in reference \cite{velten}, leading to the conclusion that slightly negative values for $\alpha$ are favoured. In our analysis below we use the set of 21 data points compiled in ref. \cite{ratra}.
\par
The SN Ia test is based on the luminosity distance, given by
\begin{eqnarray}
\mu &=& m - M = 5\log_{10} D_L\;,\\
D_L &=& \frac{c}{H_0}(1 + z)\int_0^z\frac{dz'}{E(z')}\;.
\end{eqnarray}
For the SN Ia data, we have two main problems. The first one concerns the choice of the sample. There are many different SN Ia data set, obtained with different techniques. In some cases, these different samples may give very different results. The second point is the existence of two different calibration methods: one using cosmology and which takes into account SN with high $z$ (Salt2); the other using astrophysics methods, valid for small $z$ (MLCS2k2) \cite{ioavbis}. In some case, the employment of different calibrations can lead to different results also. All this, make the SN Ia analysis very delicate. Here, however, we use the Union sample \cite{union}, calibrated by the Salt2 method. This choice is motivated by looking for a contact with previous results
in the literature, including those of reference \cite{velten}.
\par
The position of the first peak of the CMB spectrum is a more complex test. It is linked to oscillations of the baryon-photon plasma at the recombination period and is given by
\begin{equation}
l_1 = l_A(1 - \phi_1)\;.
\end{equation}
A detailed numerical analysis leading to the following fitting formula for parameters of this fundamental quantity \cite{doran, hu, doranbis, reese}:
\begin{eqnarray} \phi_1 &=& 0.267\biggr(\frac{\Omega_{r0}z_{ls}}{0.3(\Omega_{m0} + \Omega_{b0})^2}\biggl)^{0.1}\;,\\
g_1 &=& 0.0783\biggr(\omega_{b0}^{-0.238)}\biggl)\biggr(1 + 39.5\omega_{b0}^{0.763}\biggl)^{-1}\;,\\
g_2 &=& 0.560\biggr(1 + 21.1\omega_{b0}^{1.81}\biggl)^{-1}\;,\\
z_{ls} &=& 1048\biggr(1 + 0.00124\omega_{b0}^{-0.738}\biggl)\biggr(1 + g_1\omega_{m0}^{g_2}\biggl)\;,\\
l_A&=& \pi\frac{I_1}{I_2}\;,\\
I_1 &=& \int_0^{z_{ls}}\frac{dz}{E(z)}\;,\quad I_2 = \int_{z_{ls}}^\infty \frac{c_s^2(z)}{E(z)}\;,\\
c_s^2 &=& \biggr(3 + \frac{9}{4}\frac{\Omega_{b0}}{\Omega_{\gamma0}}(1 + z)\biggl)^{-1/2}\;.
\end{eqnarray}
We will use the following values for the different density parameter:
\begin{eqnarray}
\omega_{i0} = \Omega_{i0}h^2\;, \quad \Omega_{r0}h^2 = 4.116 \times 10^{-5}\;,\\ 
\Omega_{b0}h^2 = 0.02258\;,\quad \omega_{m0}h^2 = \Omega_{dm0} + \Omega_{b0}\;.
\end{eqnarray}
From observation: $l_1 = 220.08 \pm 0.7$.

The baryonic acoustic oscillations are due to the effect of oscillations in the photon-baryon plasma at the moment of the decoupling, at about $z = 1090$ (it is of course the same physics which produces the acoustic peaks structure in the CMB, but now this effect is observed in the baryonic distribution). This effect is quantified by the following expression \cite{eise}:
\begin{equation}
{\cal A} = \frac{\sqrt{\Omega_{m0}}}{E(z)^\frac{1}{3}}\biggr(\frac{1}{z_b}\int_0^{z_b}\frac{dz}{E(z)}\biggl)^\frac{2}{3}\;.
\end{equation}
In this work we use data from the WiggleZ Dark Energy Survey \cite{wiggleZ}.

The GCG behaves as dust in the past. Hence, we should identify an effective DM component in the GCG in order to use the above formulas for the analysis. There are different prescriptions in this sense in the literature. We will adopt the decomposition proposed in \cite{lu}, where the effective DM component is given by,
\begin{equation}
\label{decomposition}
\Omega_{m0} = \Omega_{b0} + \Omega_{c0}\biggr(1 - \bar A\biggl)^\frac{1}{1 + \alpha}\;.
\end{equation}

\section{Analysis and results}\label{Sec:AnandRes}

We perform a Bayesian statistic analysis in comparing the theoretical predictions with the observational data. First, for each dataset, we compute
\begin{equation}
\chi^2 = \sum_{i=1}^N\frac{(\mu_i^{th} - \mu_{ob})^2}{\sigma_i^2}\;,
\end{equation}
where $\mu_i^{th}$ is the theoretical prediction for the quantity $\mu$, whereas $\mu_i^{ob}$ is the corresponding observational datum, with an error estimation $\sigma_i$. From the chi-squared, we build the probability distribution function (PDF), as follows:
\begin{equation}\label{PDF}
P(h,\alpha,\bar A) = C\exp\biggr( - \frac{\chi^2}{2}\biggl)\;,
\end{equation}
where we assumed the data to be independent and normally distributed; $C$ is a normalization constant. As indicated, the PDF depends on three free parameters: $h$, $\alpha$ and $\bar A$, characterizing a three dimensional function. Two and one dimensional PDF can be computed integrating (marginalizing) on the remaining parameters.
\par
As already said, we will compute the $\chi^2$ separately for each dataset. The total $\chi^2$, for the set of the four observational tests, shall be the sum of the separated $\chi^2$. Accordingly to the exponential form of the PDF, in Eq.~\eqref{PDF}, the total PDF shall be the product of the single ones.
\par
The intervals considered for each of the free parameters are crucial for the final estimations. For $\bar A$, all values $0 \leq \bar A \leq 1$ are considered. For $h$ we consider both delta and constant priors i.e., for the former prior we fix values for $h$ while for the latter we integrate in the interval $0 \leq h \leq 1$.
For $\alpha$, the situation is more complex, since we can consider two possibilities: $\alpha > - 1$, in order to assure a transition from dust in the past to a cosmological constant in the future, or $\alpha > 0$, in order to assure a positive square speed of sound. Our goal here is to show that negative values for $\alpha$ are preferred. Therefore, we focus our analysis on the prior $\alpha > -1$. However, for the sake of completeness, we also carry out a statistical analysis leaving $\alpha$ free to vary to arbitrarily negative and positive values.  
\par
We evaluate the one dimensional PDF for $\bar A$ and $\alpha$ for the four independent tests listed before and combining all of them. The results are shown in \figurename{~\ref{Fig1} - \ref{Fig6}}.

\subsection{Delta prior over h}

\begin{center}
\begin{figure}[p]
\includegraphics[width=0.4\linewidth]{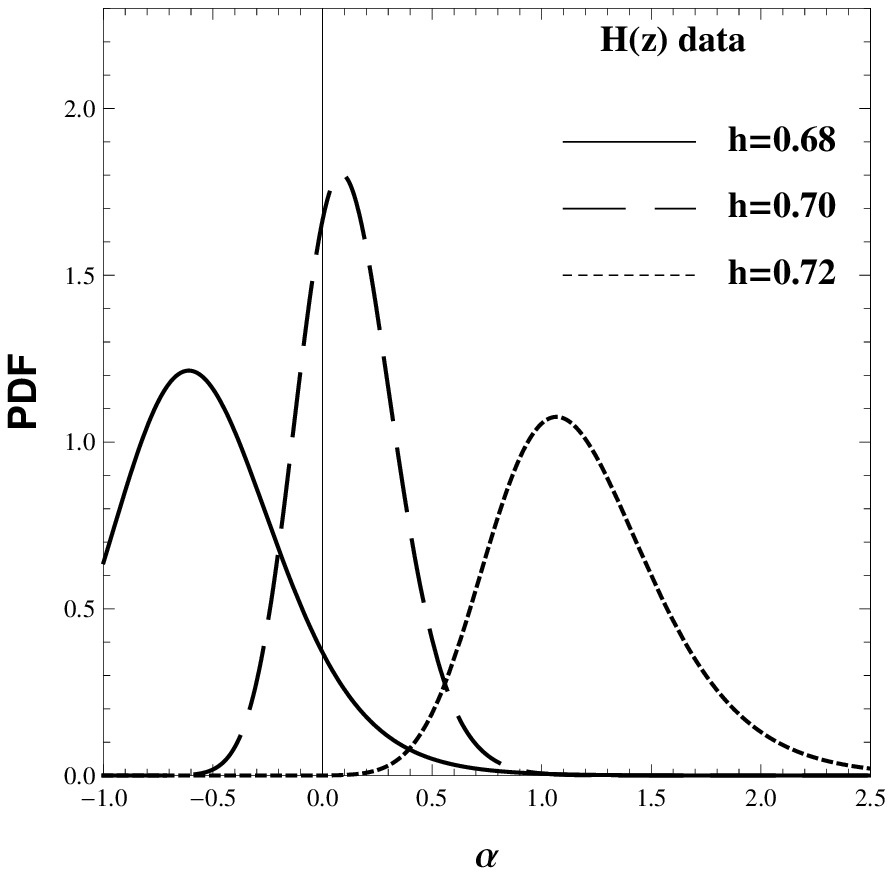}\includegraphics[width=0.4\linewidth]{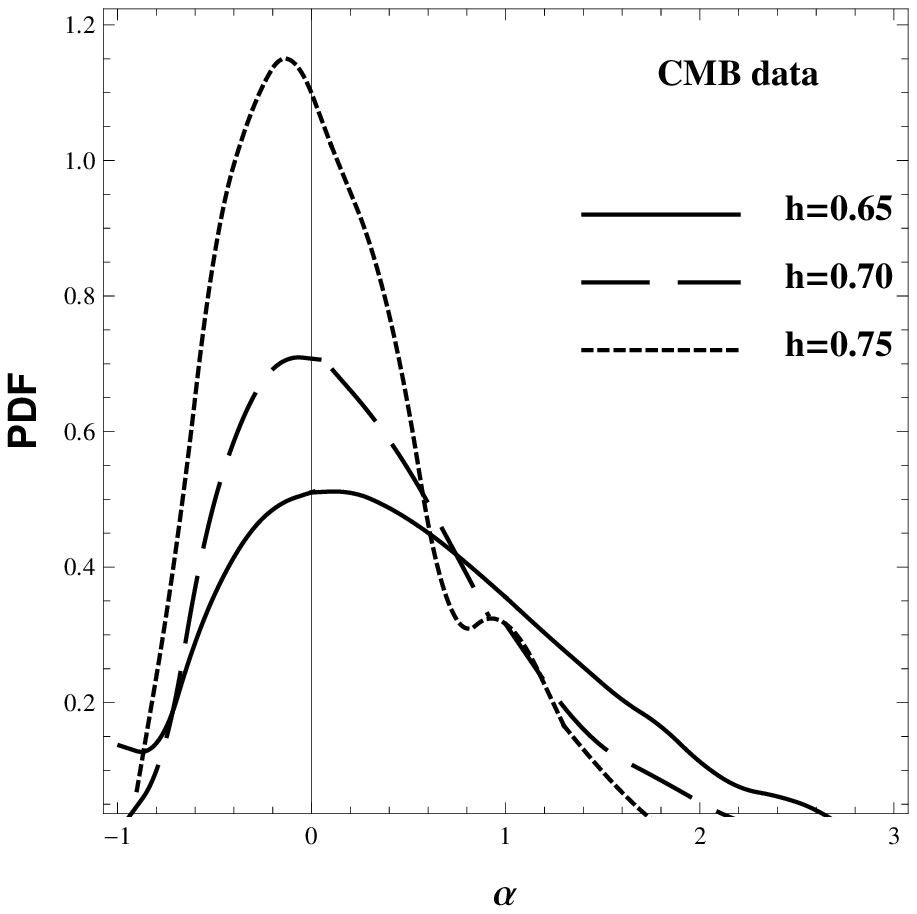}\\
\includegraphics[width=0.4\linewidth]{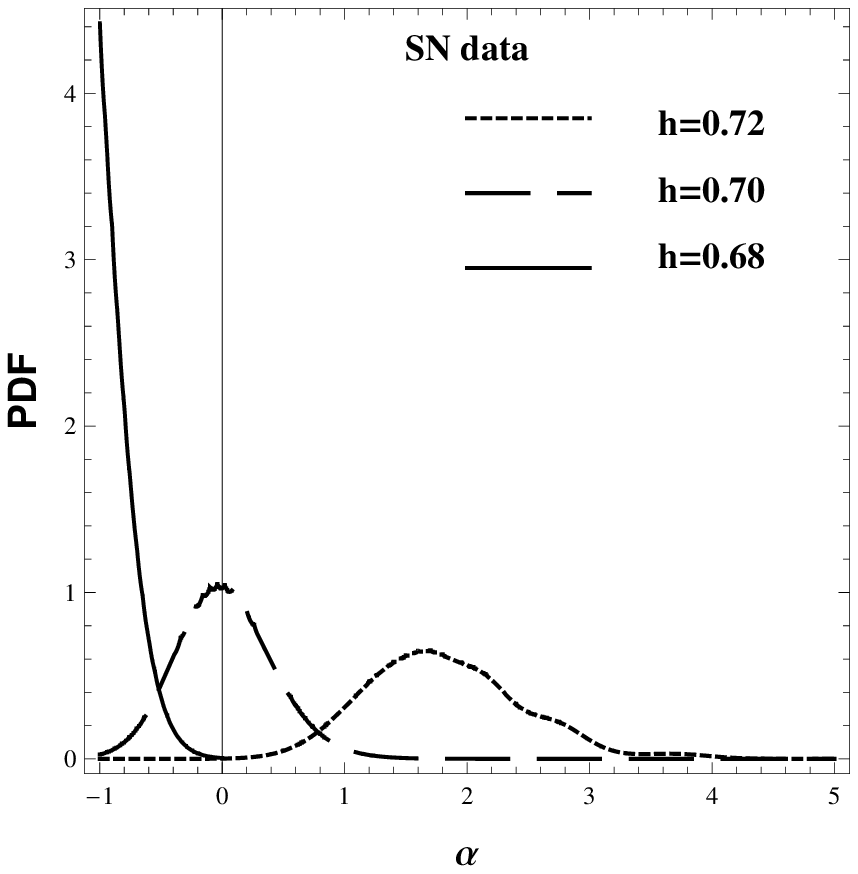}\includegraphics[width=0.4\linewidth]{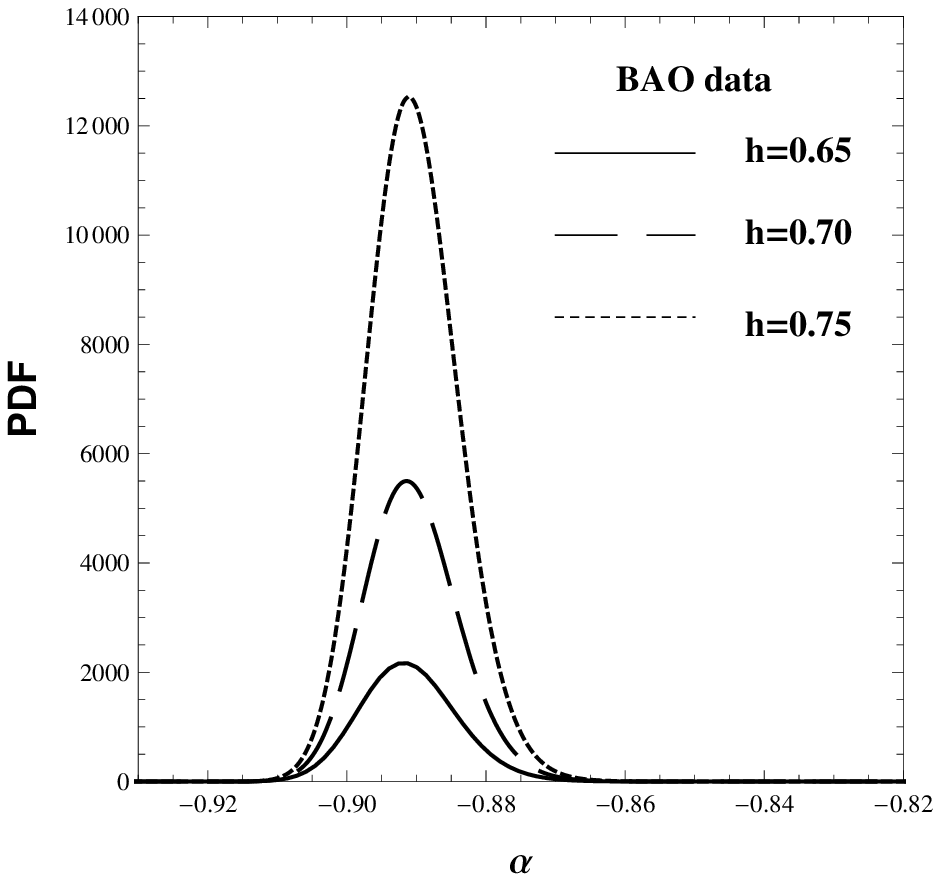}
\caption{One-dimensional PDF for the parameter $\alpha$ under the restriction $\alpha > -1$ using $H(z)$ (upper left panel), CMB (upper right panel), SNIa  (lower left panel) and BAO (lower right panel). We have used here delta priors of $h$.}
\label{Fig1}
\end{figure}
\end{center}

\begin{center}
\begin{figure}[p]
\includegraphics[width=0.4\linewidth]{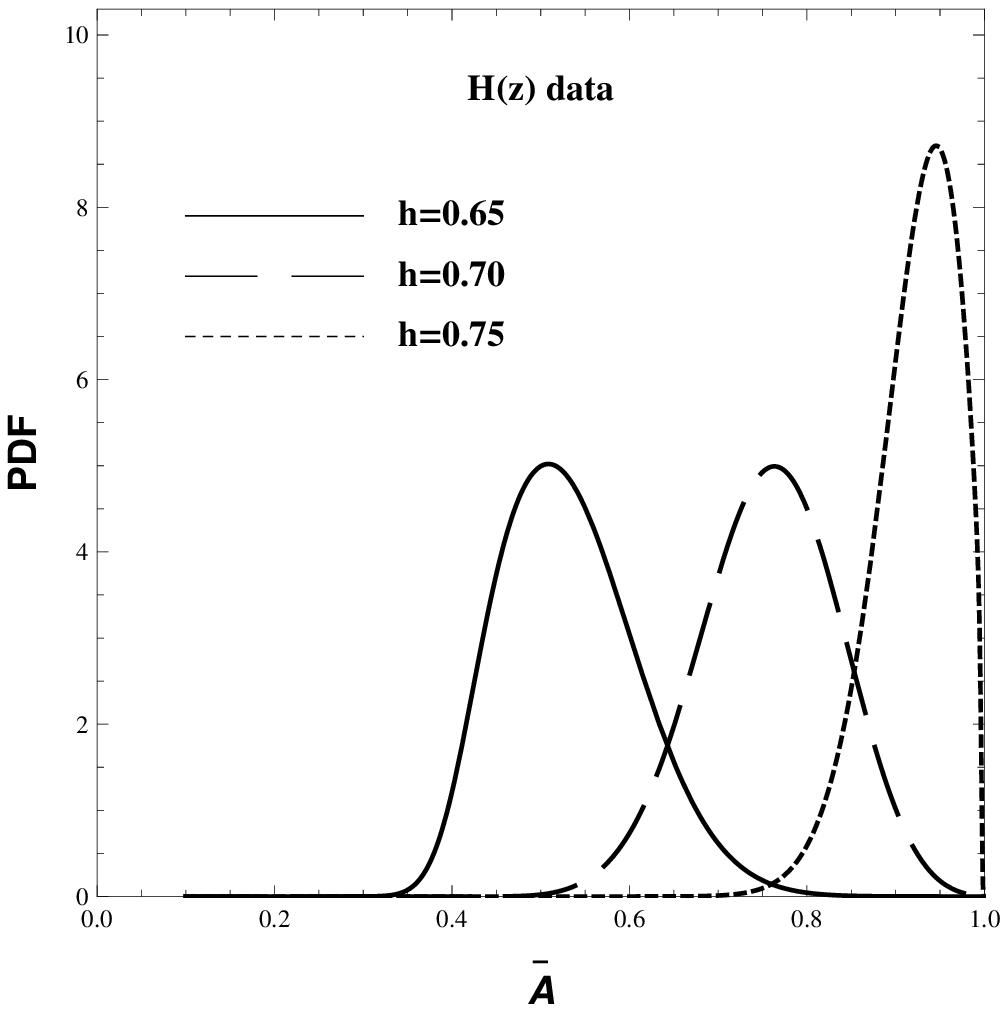}\includegraphics[width=0.4\linewidth]{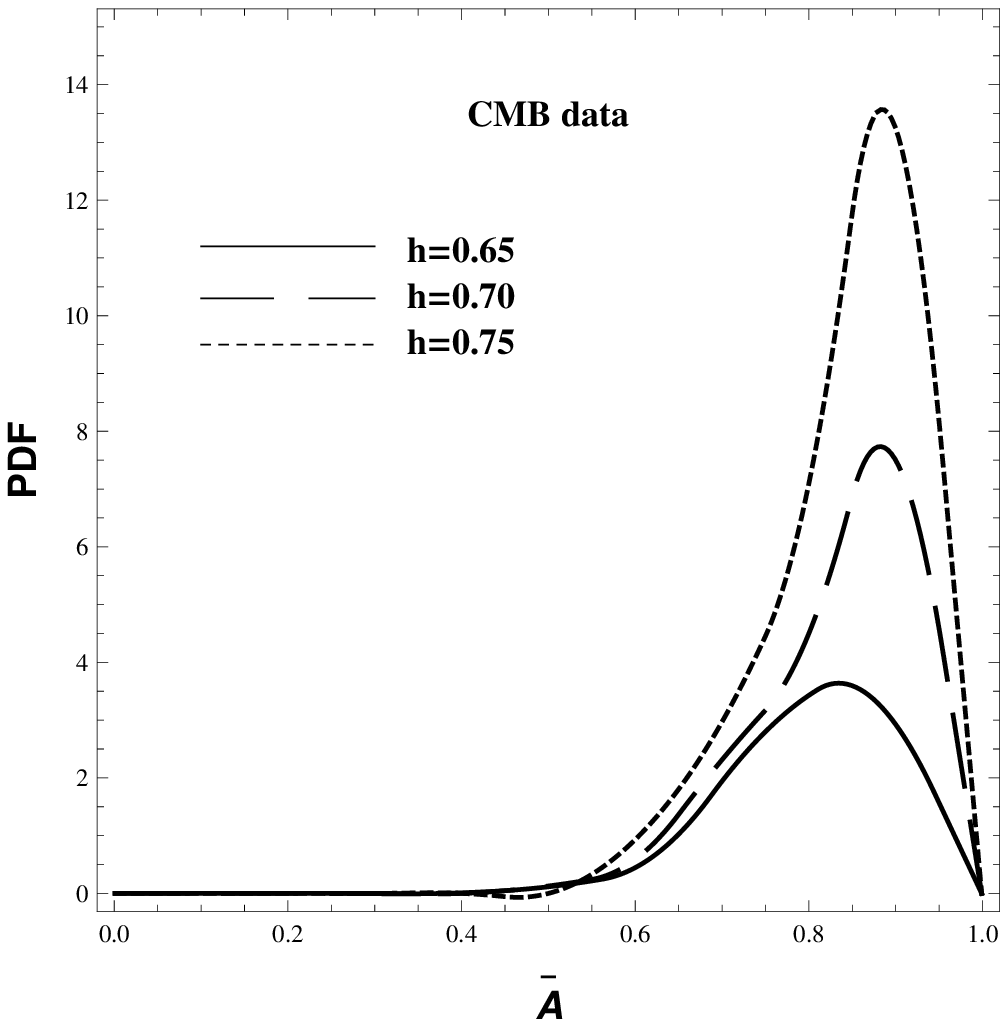}\\
\includegraphics[width=0.4\linewidth]{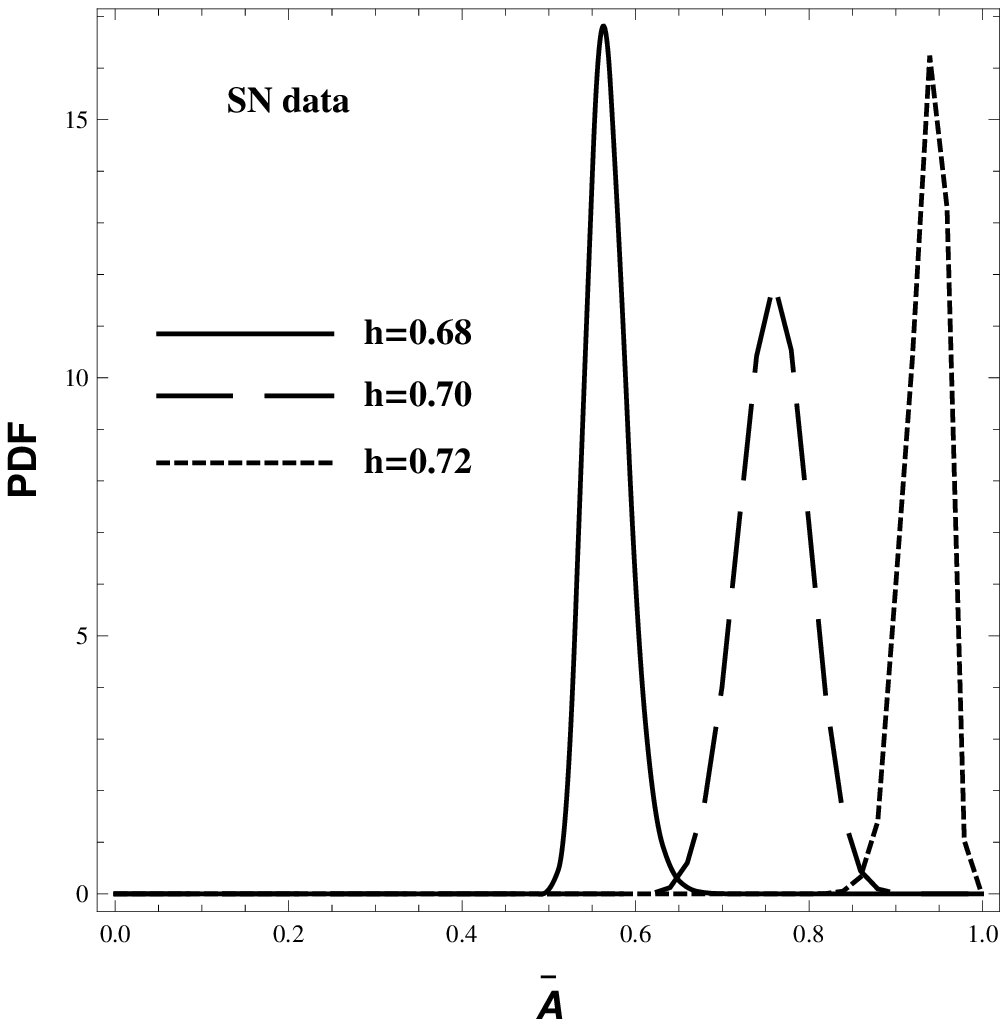}\includegraphics[width=0.4\linewidth]{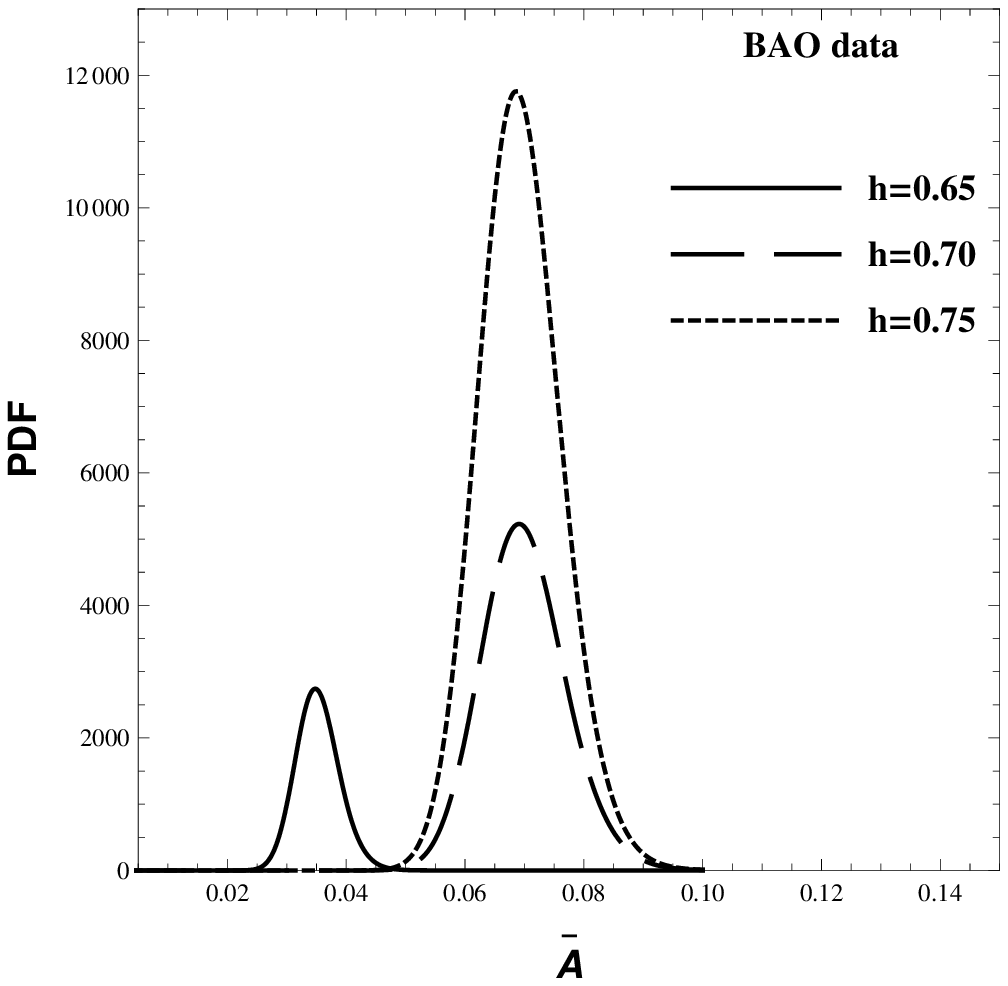}
\caption{One-dimensional PDF for the parameter $\bar{A}$ under the restriction $\alpha > -1$ using $H(z)$ (upper left panel), CMB (upper right panel), SNIa (lower left panel) and BAO (lower right panel). We have used here delta priors of $h$.}
\label{Fig2}
\end{figure}
\end{center}

In \figurename{~\ref{Fig1}, \ref{Fig2}} we show the one-dimensional probability distributions for $\alpha$ and $\bar{A}$ for specific choices of $h = h^{\star}$. The values chosen for $h^{\star}$ are shown in the four panels. Each panel corresponds to a data set. All the curves have been normalized following
\begin{equation}
{\rm PDF}(\alpha)=\frac{\int^{1}_{0}P(h^{\star},\alpha, \bar{A})\, d\bar{A}}{\int^{1}_{0}\int^{\infty}_{-1}P(h^{\star},\alpha, \bar{A}) \,d\alpha\, d\bar{A}} \hspace{0.2cm}{\rm and}\hspace{0.2cm}
{\rm PDF}(\bar{A})=\frac{\int_{-1}^{\infty}P(h^{\star},\alpha, \bar{A})\, d\alpha}{\int^{1}_{0}\int^{\infty}_{-1}P(h^{\star},\alpha, \bar{A}) \,d\alpha\, d\bar{A}}\;.
\end{equation}
Except for the H(z) data, we see that the maximum likelihoods occur for hyper-surfaces with negative values of $\alpha$.

\subsection{Flat prior over h}

For this analysis we left $h$ to vary in the range $0 < h < 1$. The final one-dimensional PDF for $\alpha$ and $\bar{A}$ are calculated formally as
\begin{equation}
{\rm PDF}(\alpha)=\frac{\int^{1}_{0}\int^{1}_{0}P(h,\alpha, \bar{A})\, dh \,d\bar{A}}{\int^{1}_{0}\int^{\infty}_{-1}\int^{1}_{0}P(h,\alpha, \bar{A})\, dh \,d\alpha \,d\bar{A}}\;\hspace{0.1cm}{\rm and}\hspace{0.1cm}{\rm PDF}(\bar A)=\frac{\int^{1}_{0}\int^{\infty}_{-1}P(h,\alpha, \bar{A})\, dh \,d\alpha}{\int^{1}_{0}\int^{\infty}_{-1}\int^{1}_{0}P(h,\alpha, \bar{A})\, dh \,d\alpha \,d\bar{A}}\;.
\end{equation}
It is crucial to restrict the interval for $\alpha$. If we consider the standard scenario for structure formation, in the GR context without entropic perturbations, it is required that $\alpha \geq 0$. However, if the only constraint is to impose an acceleration to deceleration transition, then the restriction is $\alpha > - 1$, since for $\alpha < - 1$ the universe accelerates in the past, and it is decelerating today. Alternatively, we may leave the value of $\alpha$ free in order to test the consistency of the background tests in the context of the GCG model. In what follows, we will consider two cases: $\alpha > - 1$ and $\alpha$ free.
\par
For both the cases, the general behaviour follows similar features, that can be summarized as follows. The $H(z)$ and the baryonic acoustic oscillations tests predict a maximum for $\alpha$ slightly negative, while the position of the first peak indicates a maximum for a small positive value of $\alpha$. For these three tests the PDF decreases as the value $\alpha = - 1$ is approached. Only for the SNIa test the PDF for $\alpha$ can be significant for $\alpha < - 1$.
It is important to stress that, for the baryonic acoustic oscillations and for the CMB first peak, the PDF becomes essentially zero for $\alpha < - 1$. This is due to the decomposition into a "dark matter" component employed in Eq.~(\ref{decomposition}): for $\alpha < - 1$, the behaviour for "dark matter" is reversed, due to the change of sign of the exponent in Eq.~(\ref{decomposition}). 
\par
\begin{center}
\begin{figure}[p]
\includegraphics[width=0.3\linewidth]{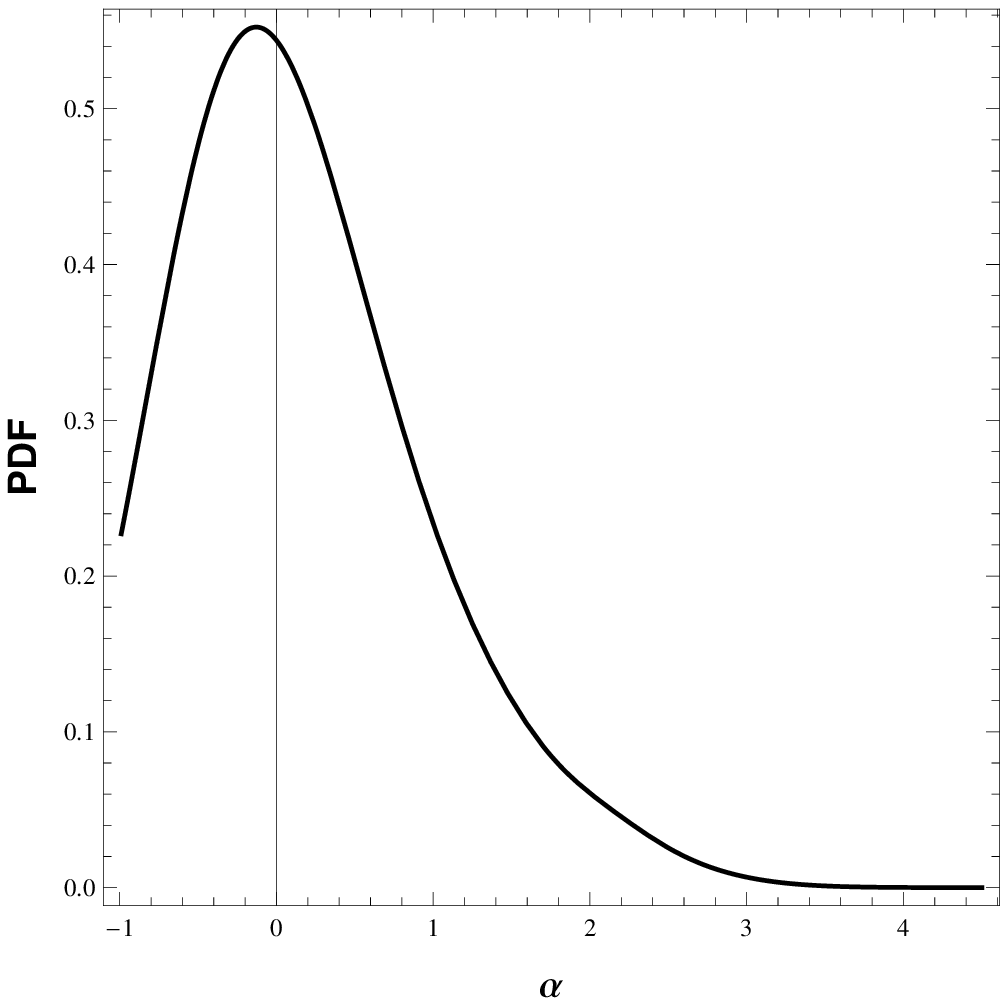}\includegraphics[width=0.3\linewidth]{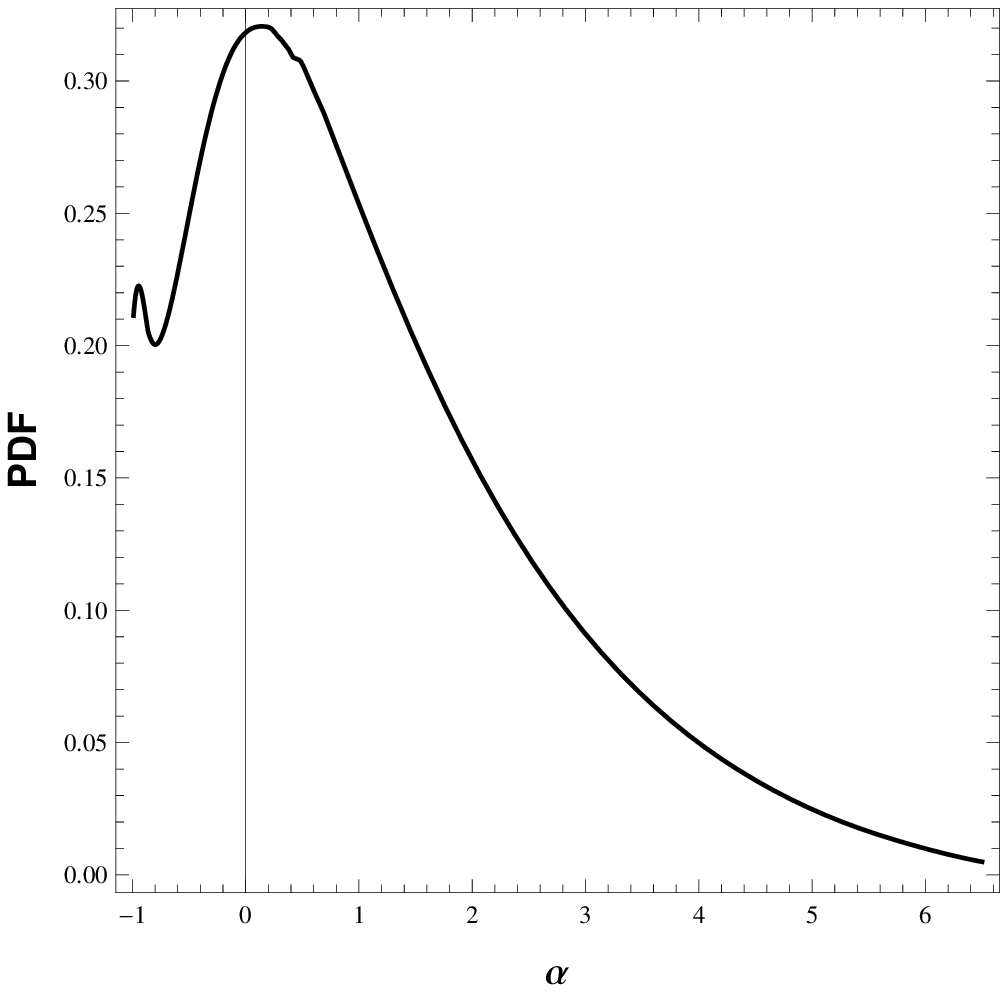}\includegraphics[width=0.3\linewidth]{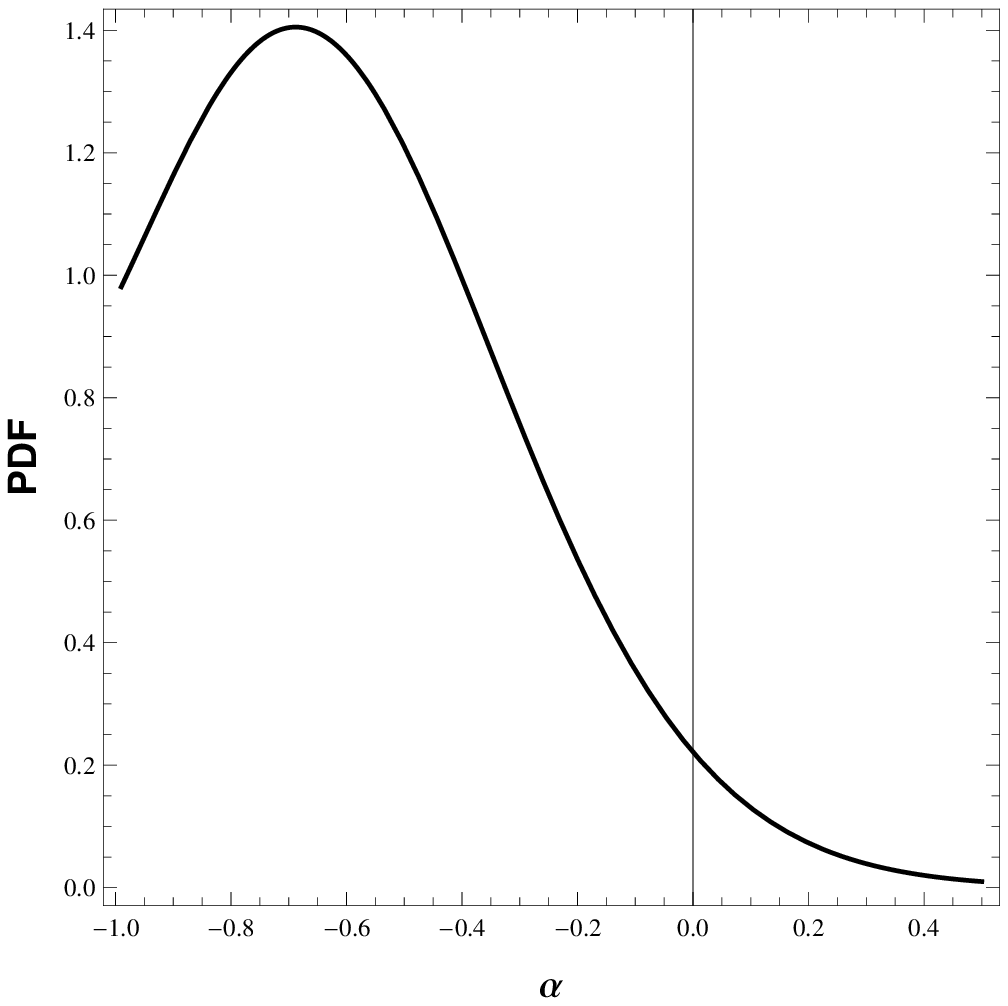}\\
\includegraphics[width=0.3\linewidth]{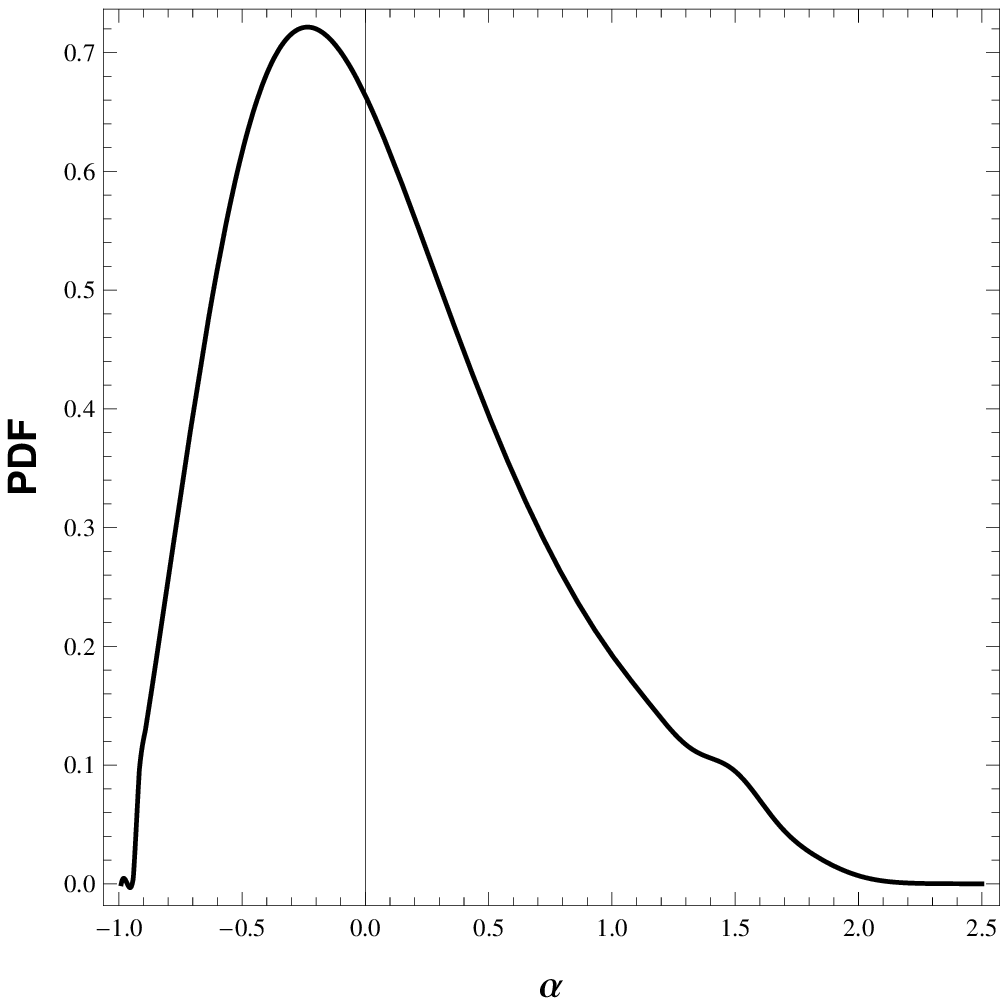}\includegraphics[width=0.3\linewidth]{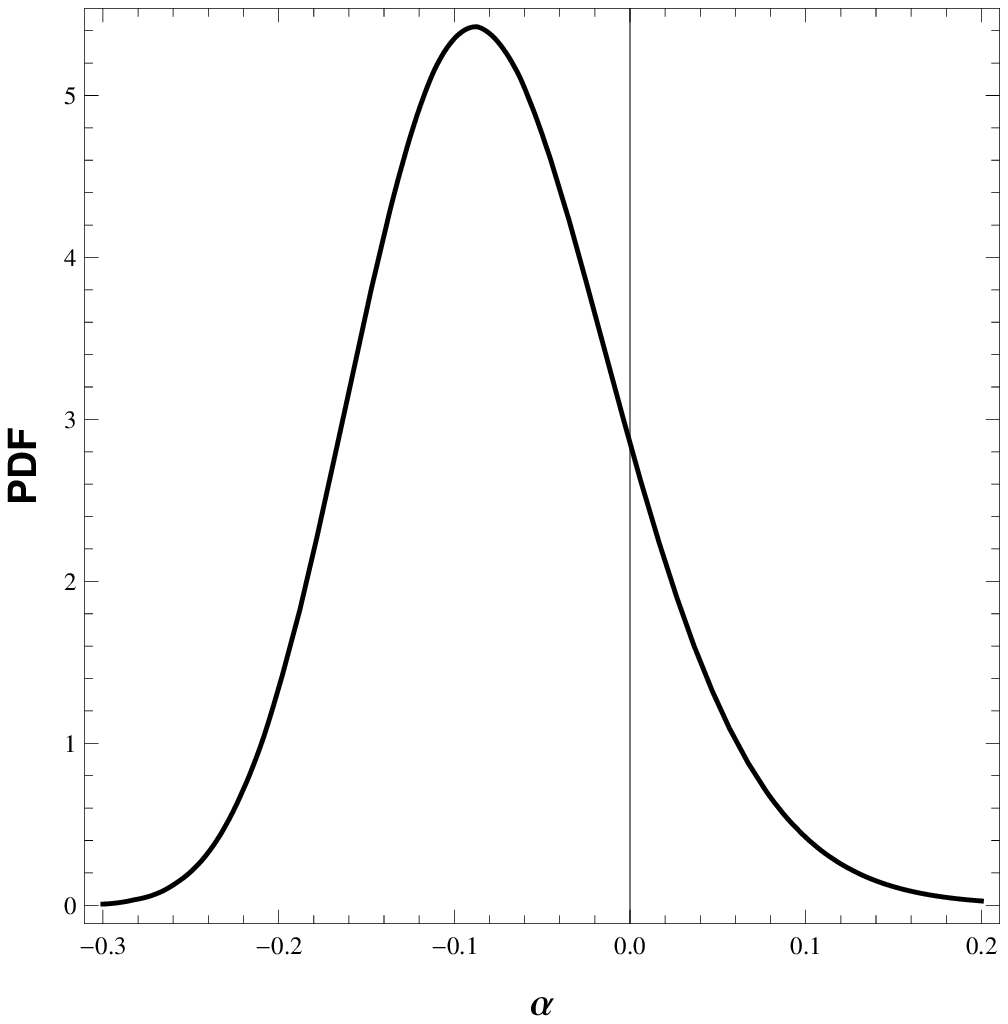}
\caption{One-dimensional PDF for the parameter $\alpha$ under the restriction $\alpha > -1$ using $H(z)$ (upper left panel), CMB (upper central panel), SNIa (upper right panel), BAO (lower left panel) and the combination of these four tests (lower right panel). We have used a flat prior for $h$ and marginalized over it.}
\label{Fig3}
\end{figure}
\end{center}
\begin{center}
\begin{figure}[p]
\includegraphics[width=0.3\linewidth]{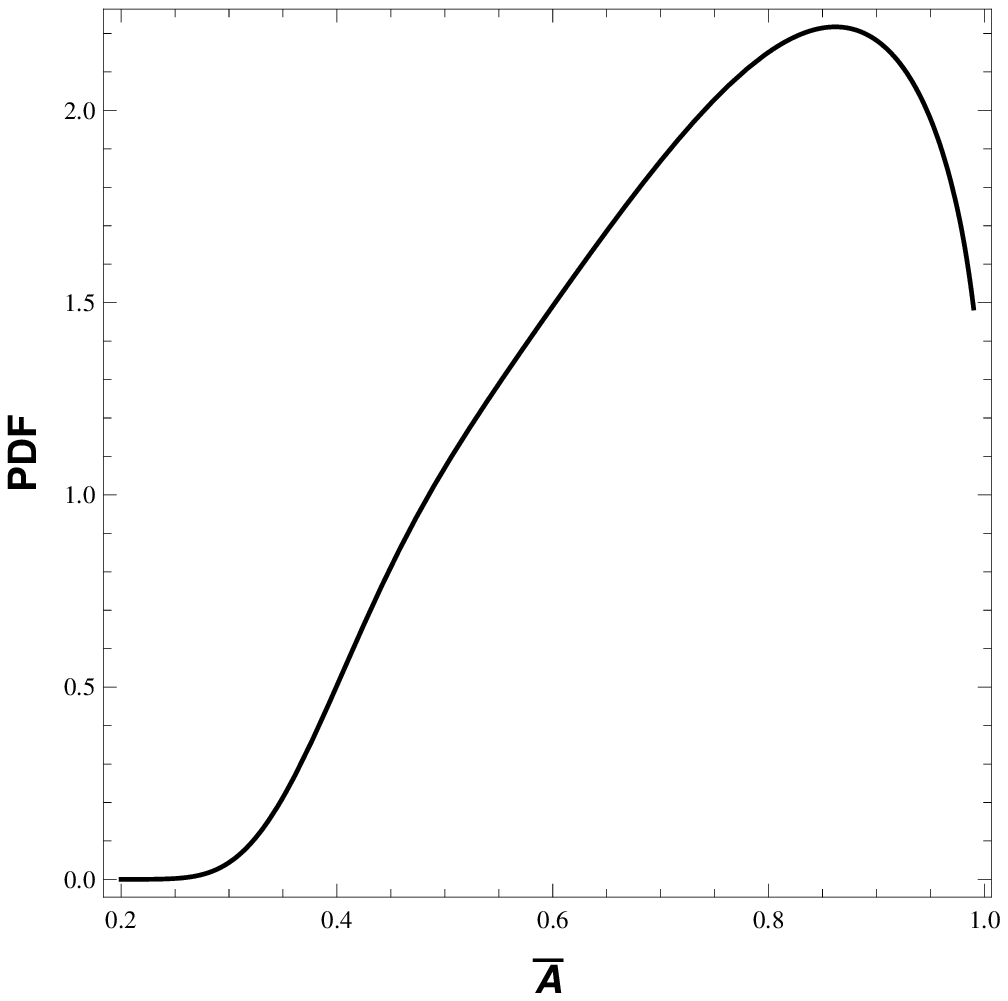}\includegraphics[width=0.3\linewidth]{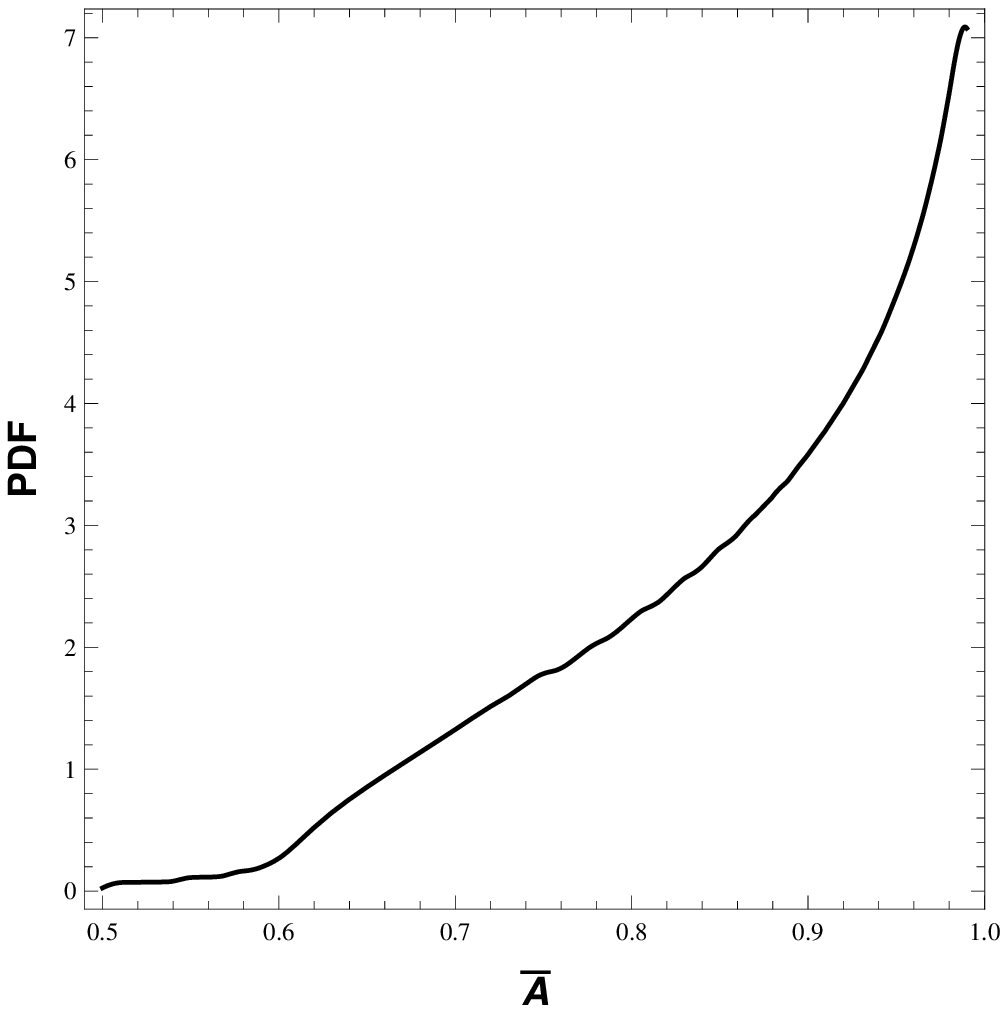}\includegraphics[width=0.3\linewidth]{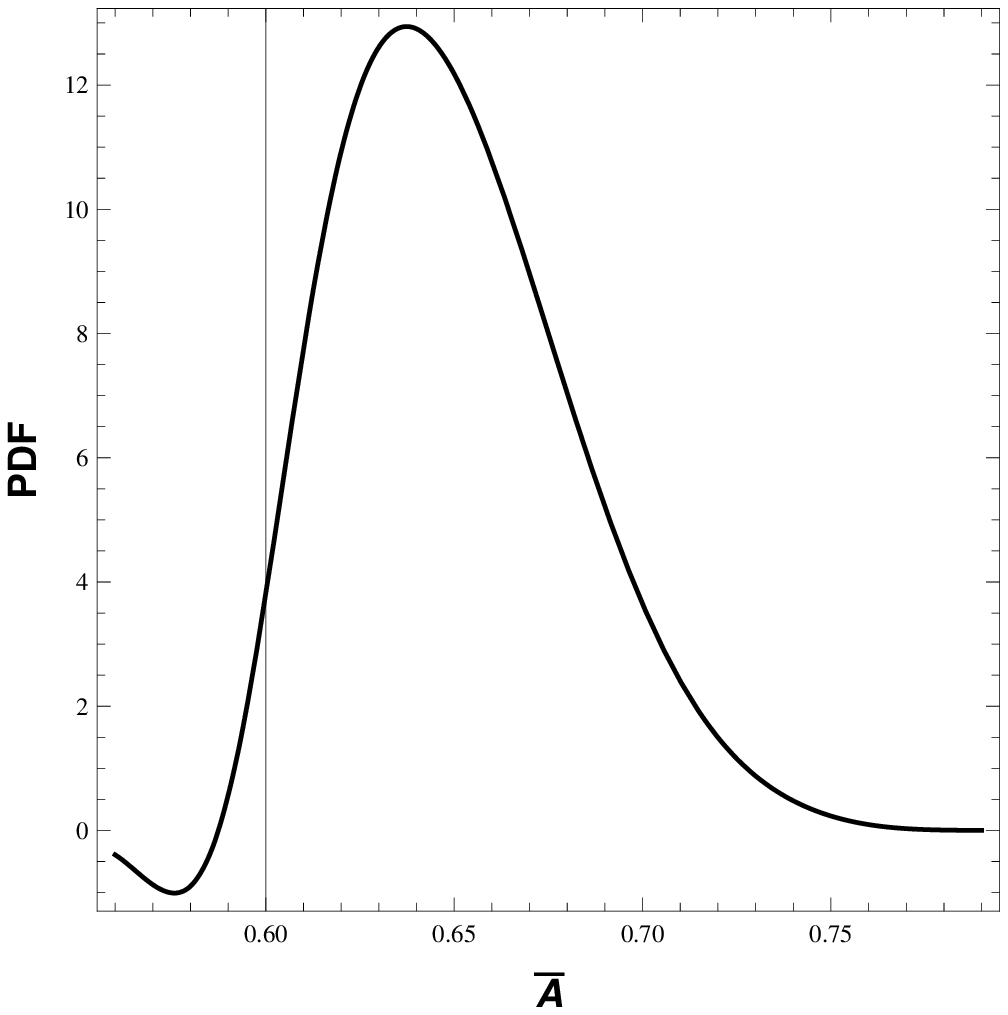}\\
\includegraphics[width=0.3\linewidth]{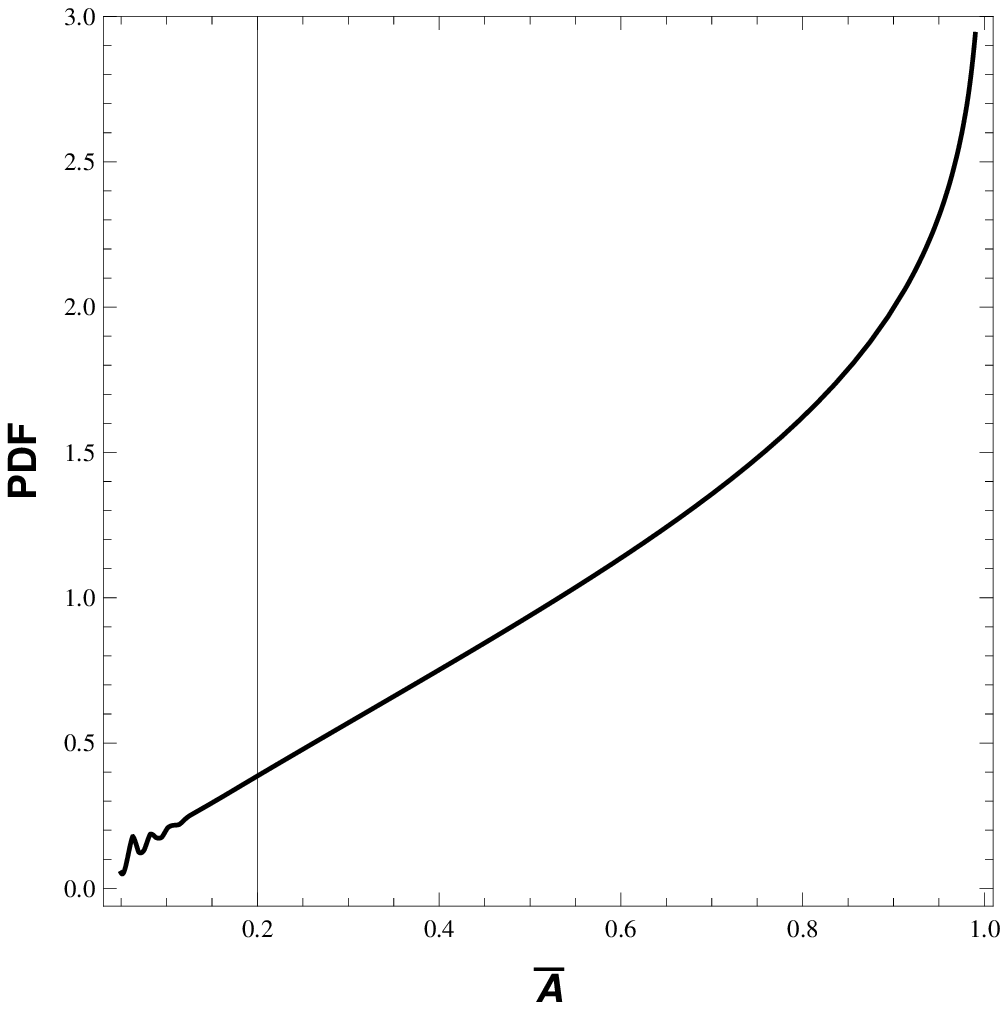}\includegraphics[width=0.3\linewidth]{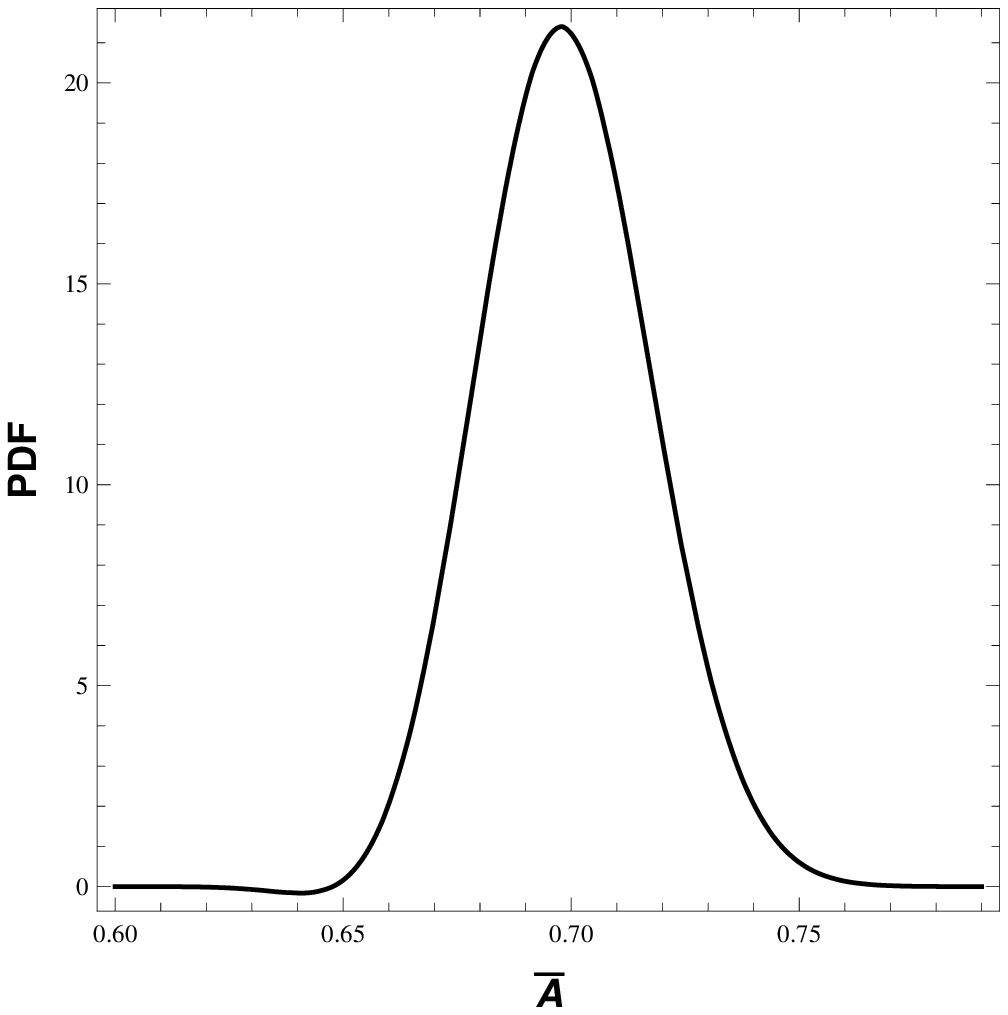}\includegraphics[width=0.3\linewidth]{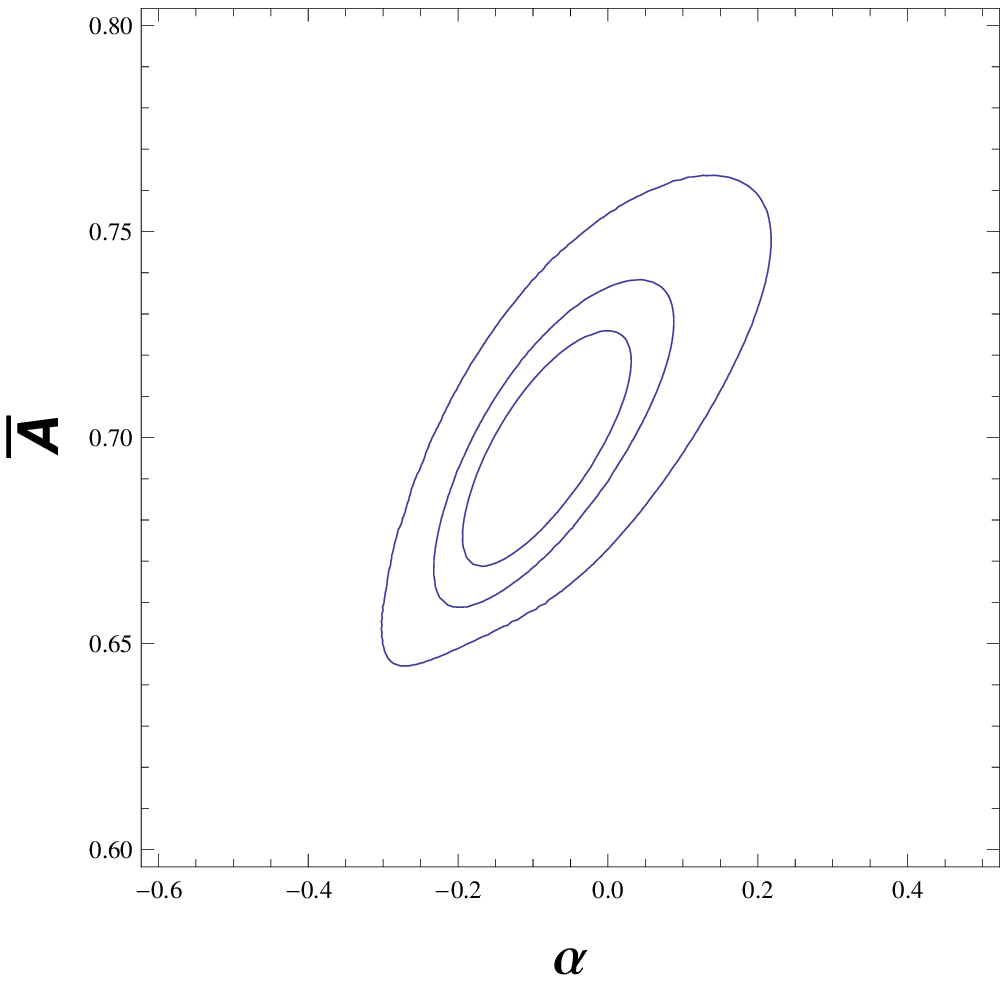}
\caption{One-dimensional PDF for the parameter $\bar A$ under the restriction $\alpha > -1$ using $H(z)$ (upper left panel), CMB (upper central panel), SNIa (upper right panel), BAO (lower left panel) and the combination of these four tests (lower central panel). In the lower right panel, we present the contour plots (1, 2, 3 $\sigma$) for the total PDF in the ($\alpha$, $\bar{A}$) plane. We have used a flat prior for $h$ and marginalized over it.}
\label{Fig4}
\end{figure}
\end{center}
\begin{center}
\begin{figure}[p]
\includegraphics[width=0.3\linewidth]{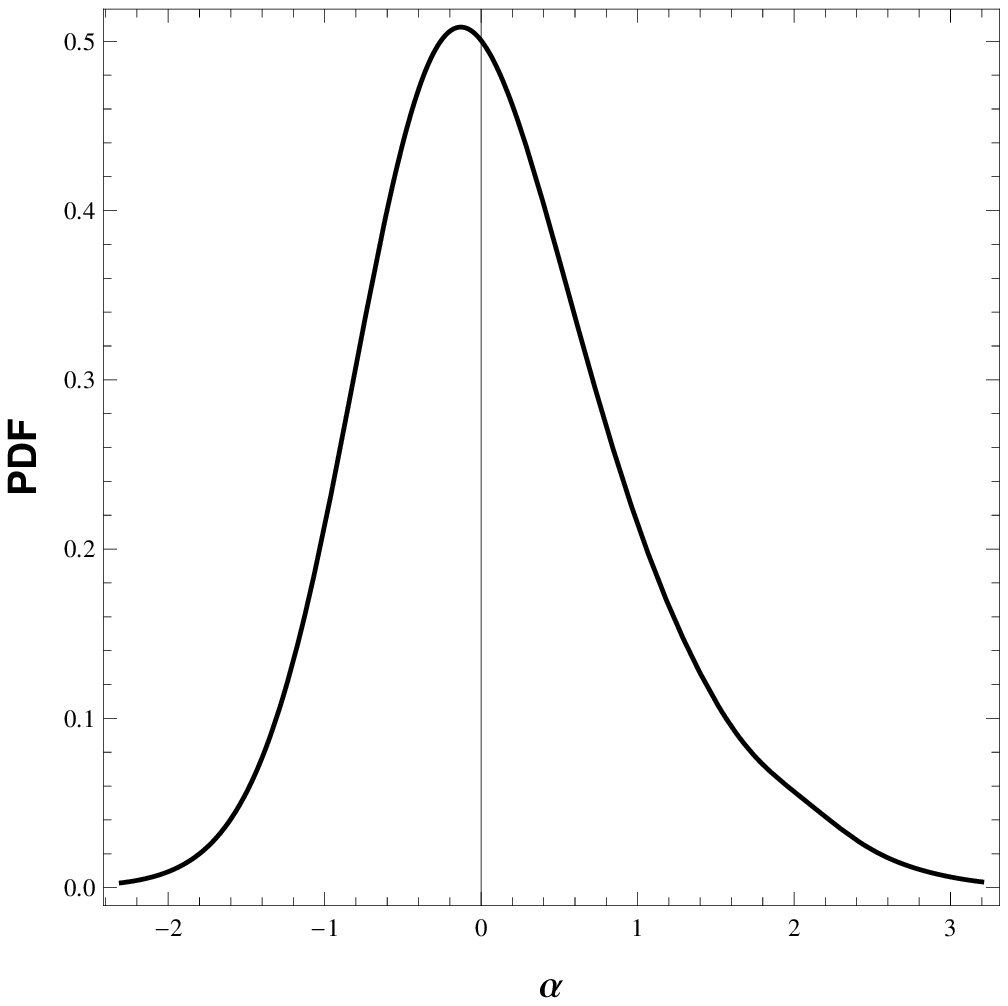}\includegraphics[width=0.3\linewidth]{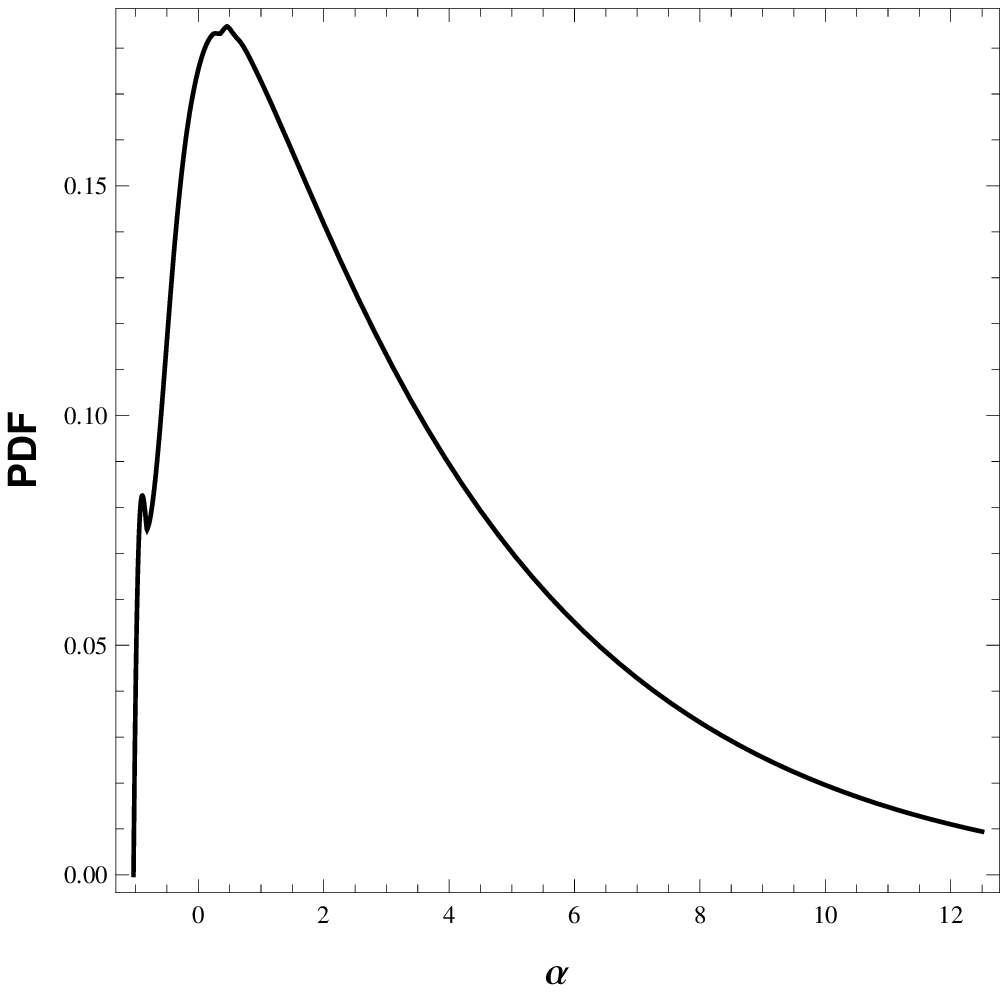}\includegraphics[width=0.3\linewidth]{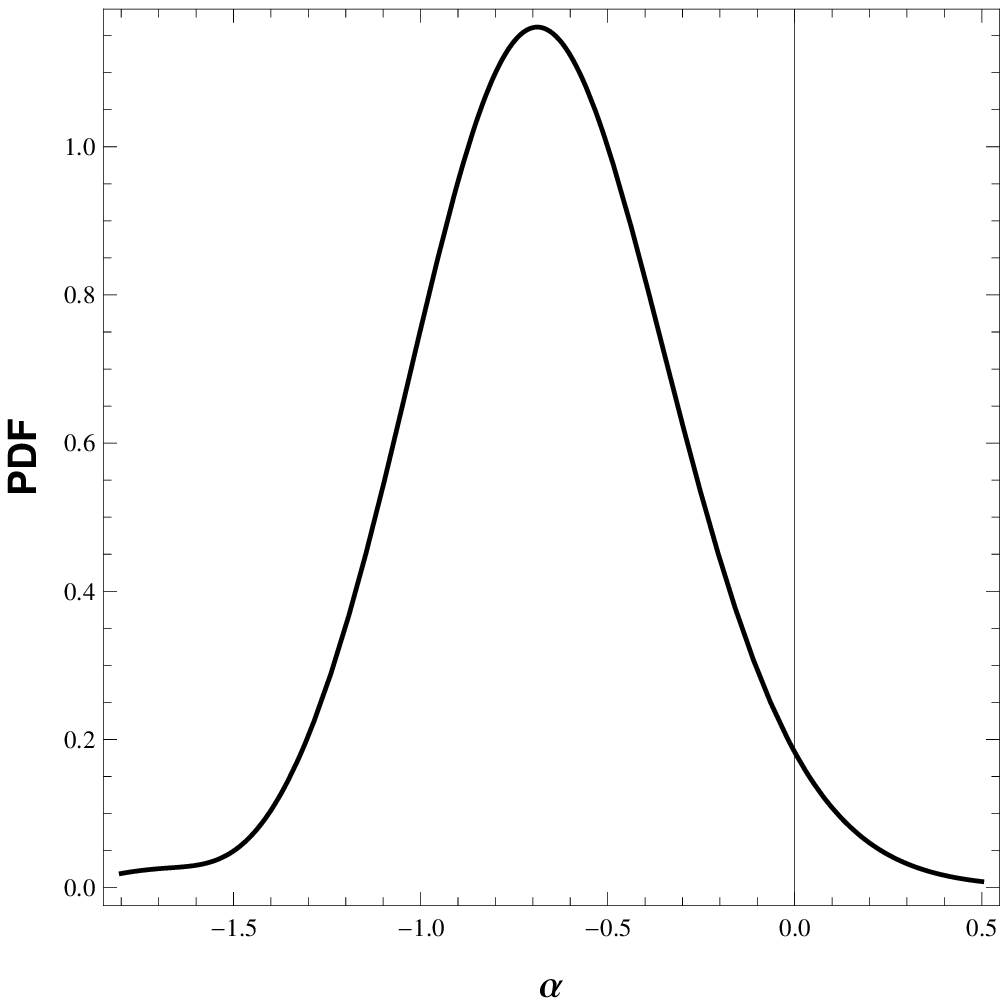}\\
\includegraphics[width=0.3\linewidth]{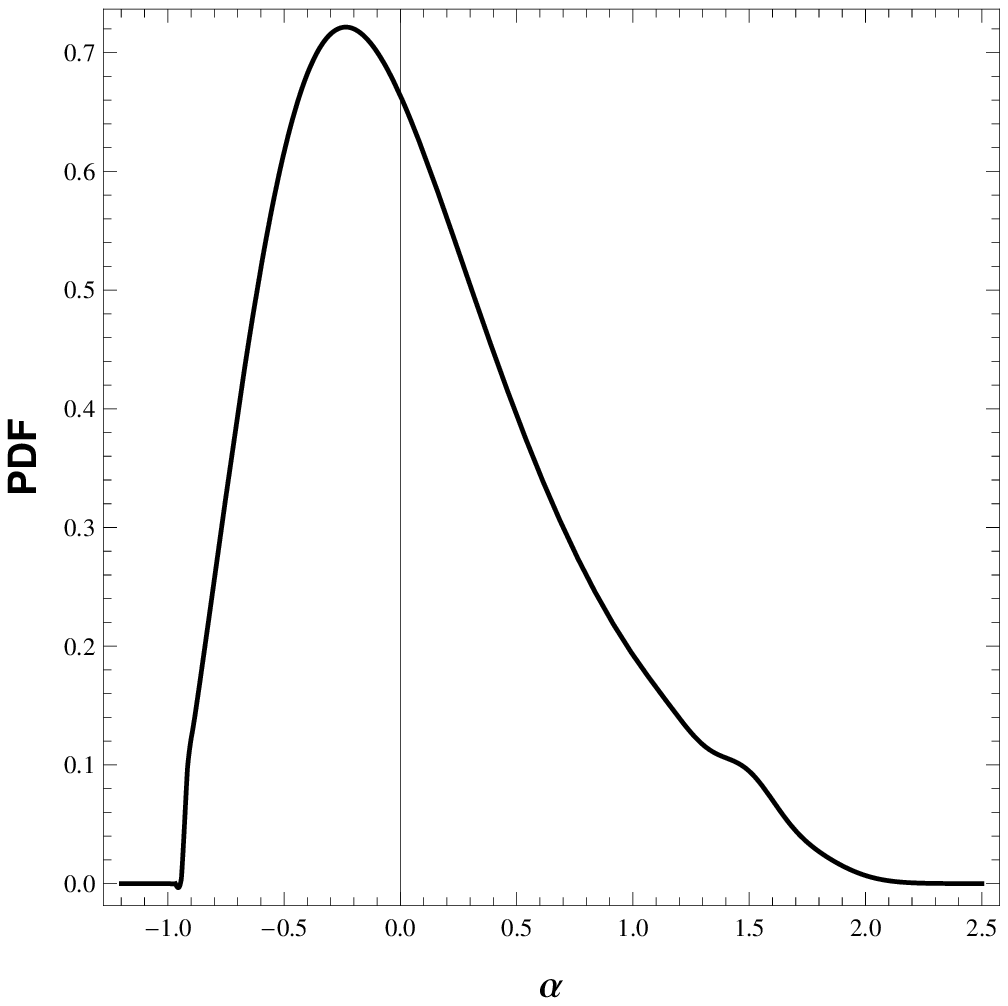}\includegraphics[width=0.3\linewidth]{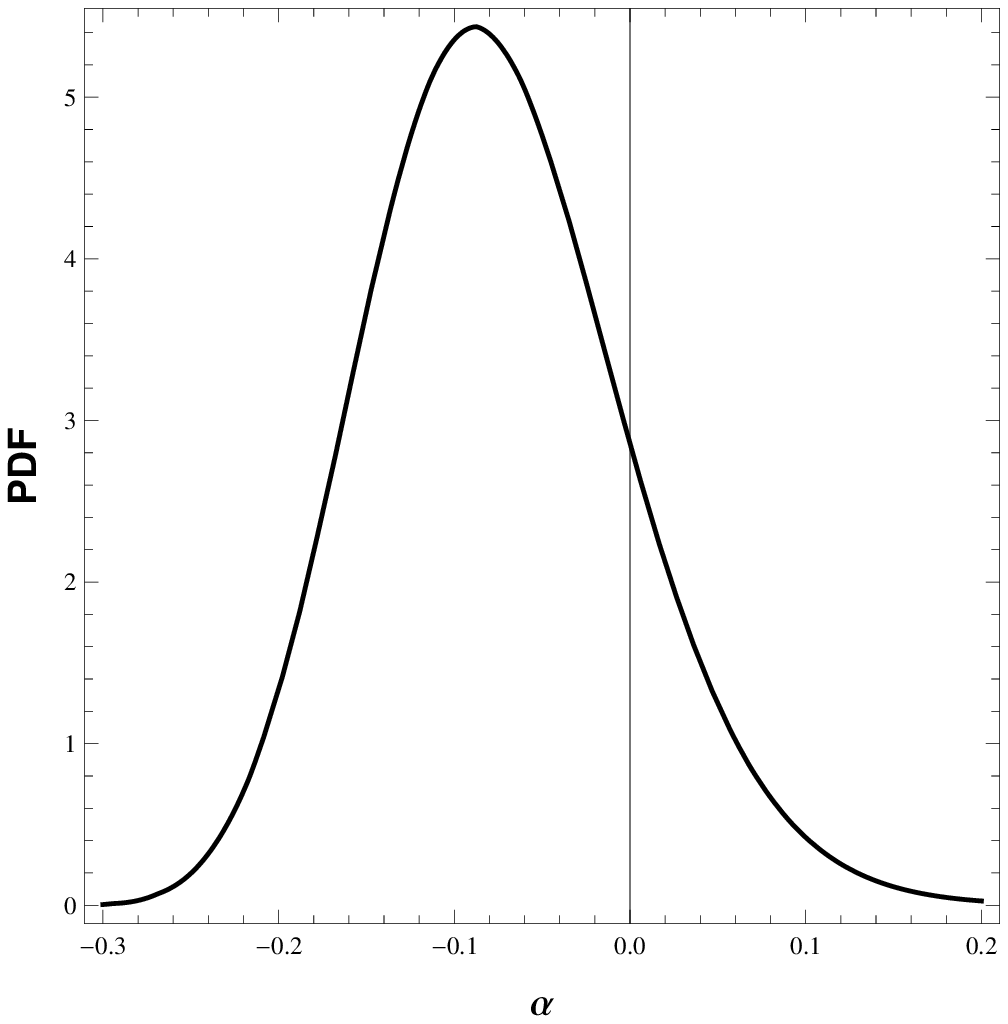}
\caption{One-dimensional PDF for the parameter $\alpha$, with no restriction, using $H(z)$ (upper left panel), CMB (upper central panel), SNIa (upper right panel), BAO (lower left panel) and the combination of these four tests (lower right panel). We have used a flat prior for $h$ and marginalized over it.}
\label{Fig5}
\end{figure}
\end{center}
\begin{center}
\begin{figure}[p]
\includegraphics[width=0.3\linewidth]{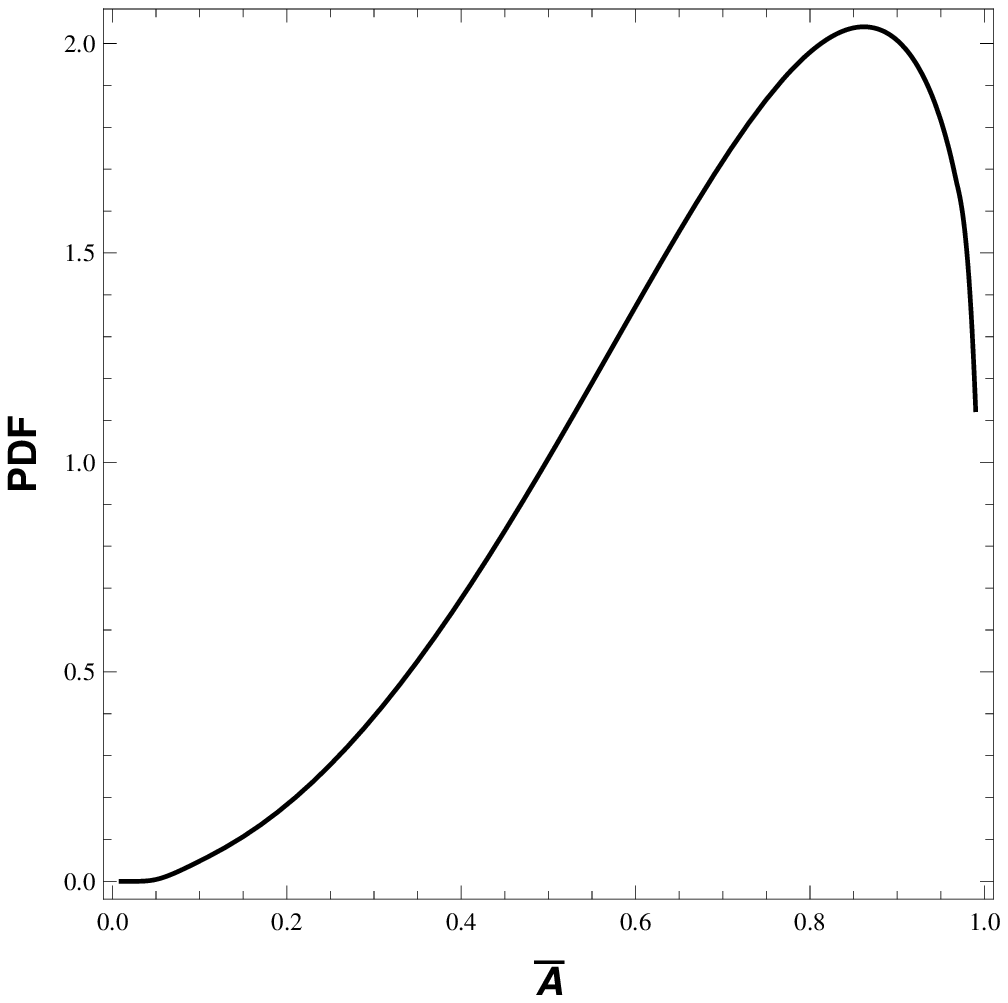}\includegraphics[width=0.3\linewidth]{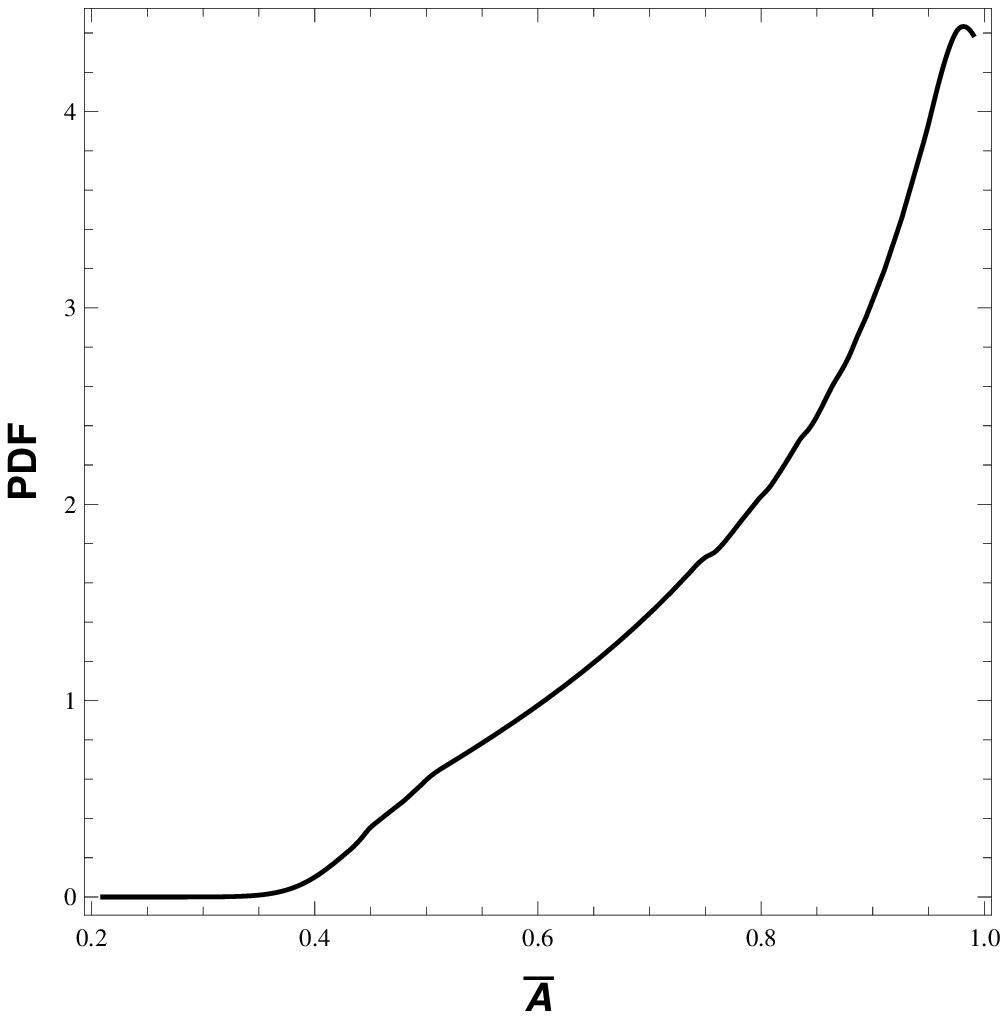}\includegraphics[width=0.3\linewidth]{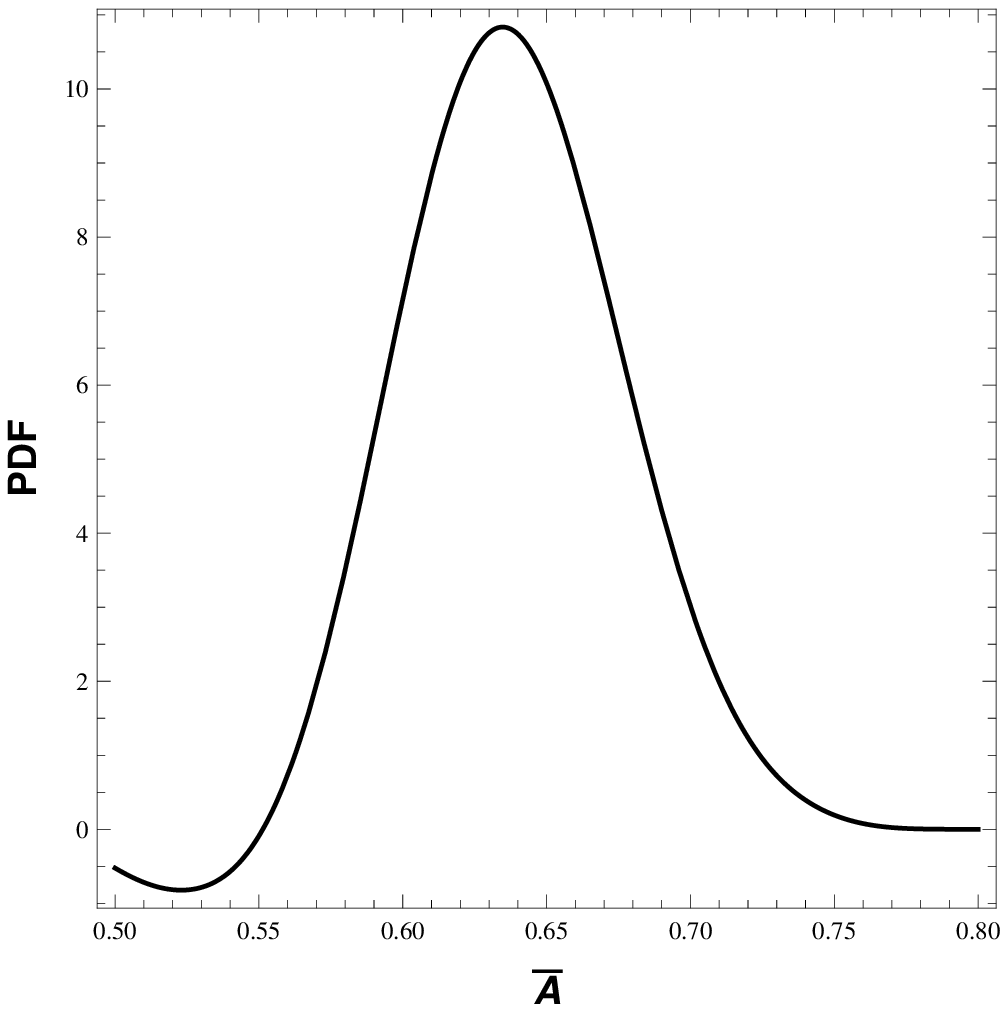}\\
\includegraphics[width=0.3\linewidth]{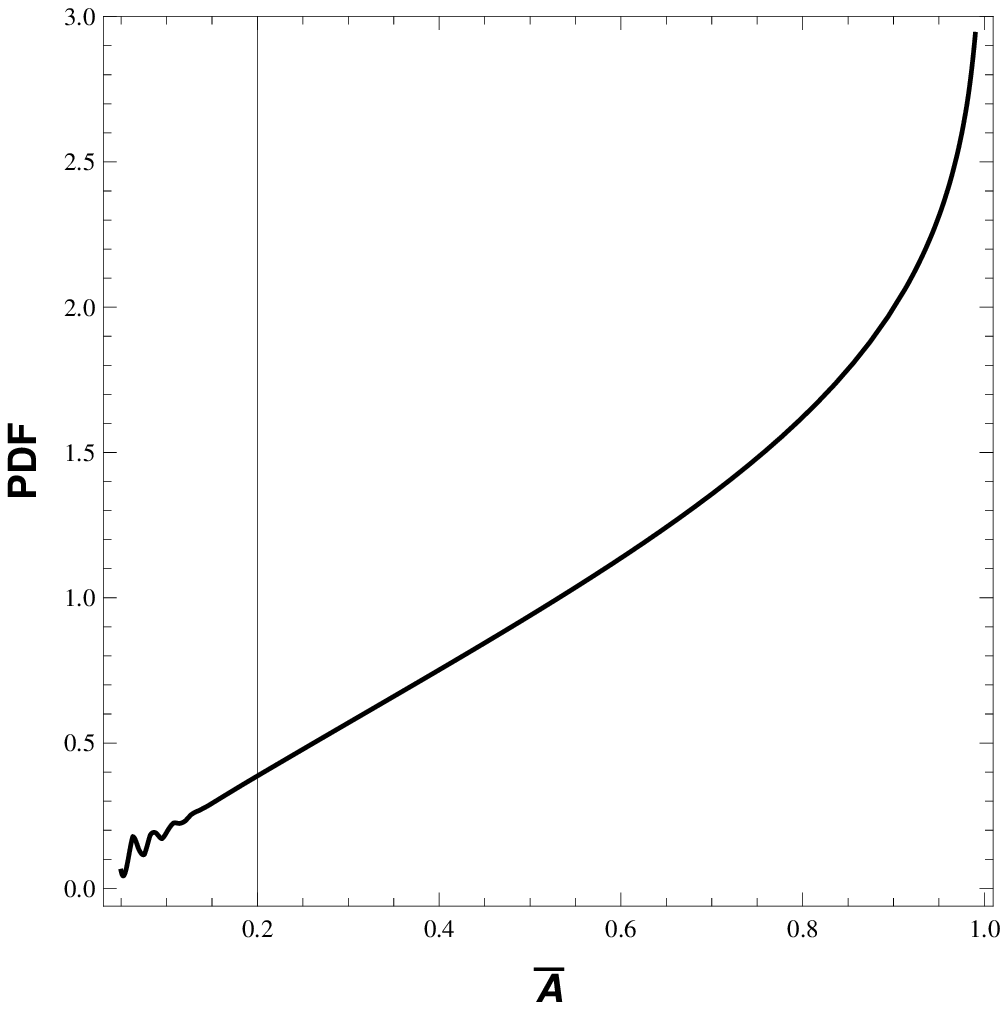}\includegraphics[width=0.3\linewidth]{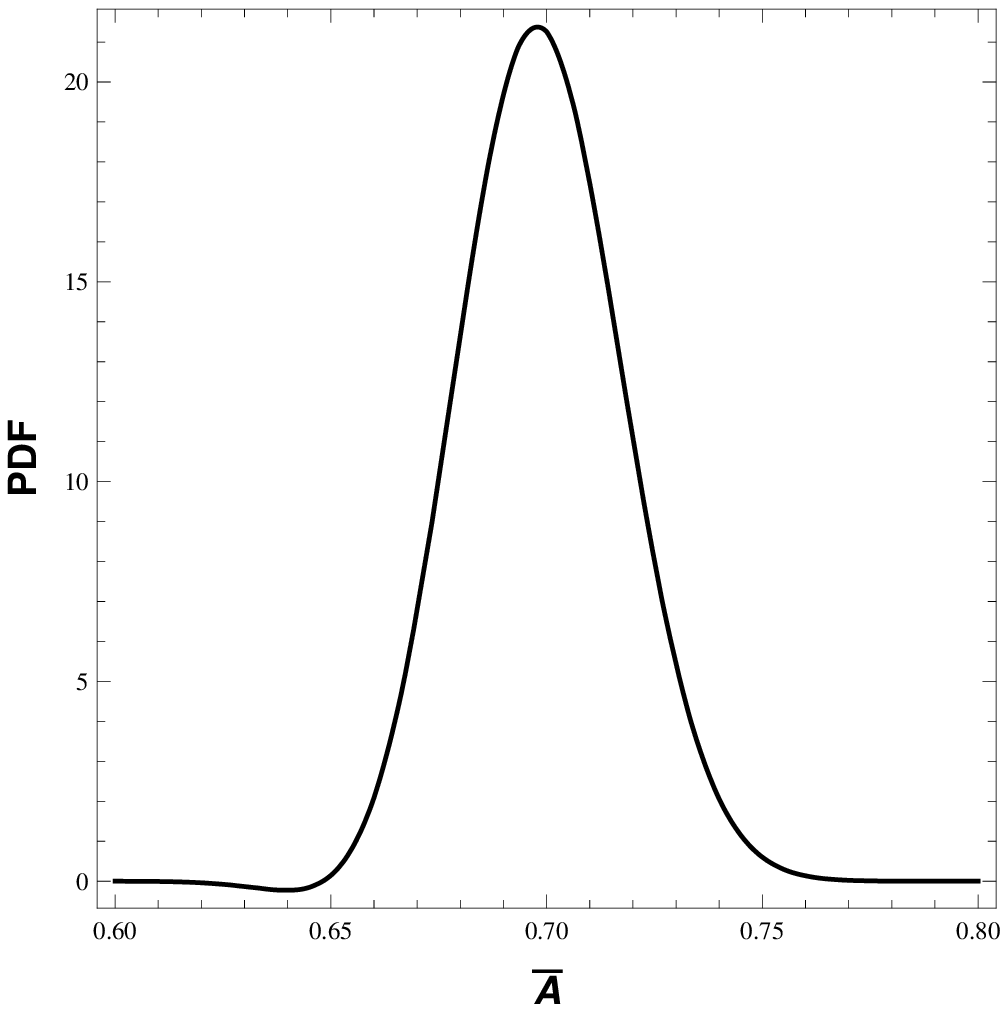}\includegraphics[width=0.3\linewidth]{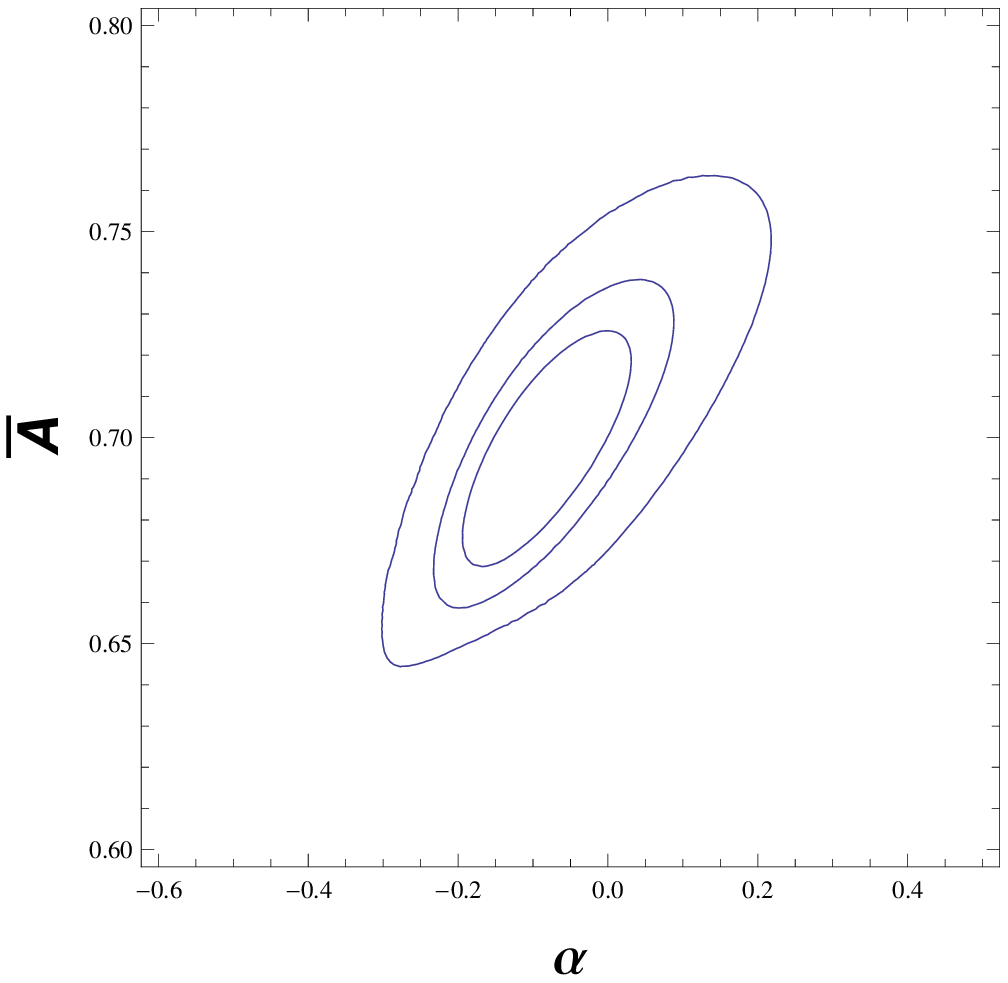}
\caption{One-dimensional PDF for the parameter $\bar A$, with no restriction on $\alpha$, using $H(z)$ (upper left panel), CMB (upper central panel), SNIa (upper right panel), BAO (lower left panel) and the combination of these four tests (lower central panel). In the lower right panel, we present the contour plots (1, 2, 3 $\sigma$) for the total PDF in the ($\alpha$, $\bar{A}$). We have used a flat prior for $h$ and marginalized over it.}
\label{Fig6}
\end{figure}
\end{center}

We show the results of the analysis in \figurename{~\ref{Fig3}} and \figurename{~\ref{Fig4}}, for $\alpha > - 1$, and in \figurename{~\ref{Fig5}} and \figurename{~\ref{Fig6}}, for $\alpha$ free. The $2\sigma$ estimation for $\alpha$ in the first case is $\alpha = - 0.087^{+0.159}_{-0.135}$, while in the second case is $\alpha = - 0.089^{+0.159}_{-0.130}$. This result shows that $\alpha > - 1$ is a good prior. Moreover, we find that $\alpha < 0$ with 85\% confidence.

\newpage

\section{A Chaplygin gas fluid description in Rastall's theory and some possible indications for its healing}\label{Sec:Healing}

Some ideas on how to avoid a possible ``tension'' and then saving the Chaplygin gas appeared recently and concern a modified theory of gravity: Rastall's theory \cite{Rastall:1973nw}. In the latter the Einstein-Hilbert action still holds and the modification resides in the conservation law of the matter stress-energy momentum, i.e.
\begin{equation}
\label{mot1}
{T^{\mu\nu}}_{;\mu} = \kappa R^{;\nu}\;,
\end{equation}
where $\kappa$ is a parameter and $R$ is the Ricci scalar curvature. The above modification can be interpreted in various ways. To us the most interesting and significant one is that Rastall's idea may be viewed as a kind of semi-classical formulation of quantum phenomena, which we expect to appear when the curvature (which enters as $R$ on the right hand side of Eq.~\eqref{mot1}) becomes important. Of course, one could choose other scalars rather than $R$ in order to represent the Riemann tensor, but perhaps $R$ is the most natural choice.

Since Eq.~\eqref{mot1} must fit into Bianchi identities, one can find the following modified Einstein equations
\begin{eqnarray}
\label{eq1R} R_{\mu\nu} - \frac{1}{2}g_{\mu\nu}R &=& 8\pi G\left(T_{\mu\nu} - \frac{\gamma- 1}{2}g_{\mu\nu}T\right)\;,\\
\label{eq2R} {T^{\mu\nu}}_{;\mu} &=& \frac{\gamma - 1}{2}T^{;\nu}\;.
\end{eqnarray}
Then, it is clear that, when assuming a FLRW background, Friedmann equation shall be modified with terms proportional to $\gamma$.

We choose, for our discussion in the present section, a single-fluid component with density $\rho$ and pressure $p$. In Rastall's theory, Friedmann equation becomes
\begin{equation}\label{friedeqrast}
 H^2 = \frac{8\pi G}{3}\rho\left[\frac{3 - \gamma}{2} + \frac{3w}{2}\left(\gamma - 1\right)\right]\;,
\end{equation}
where note that, differently from GR, now the pressure contributes to the expansion (via the equation of state parameter $w$). Moreover, the continuity equation becomes:
\begin{equation}
 \dot\rho + 3H\left(\rho + p\right) = \frac{\gamma - 1}{2}\left(\dot\rho - 3\dot p\right)\;,
\end{equation}
or, trading the cosmic time for the scale factor, and rearranging the derivatives:
\begin{equation}\label{enconsrast}
 \frac{3 - \gamma}{2}\frac{d\rho}{da} + \frac{3}{a}\left(\rho + p\right) = -3\frac{\gamma - 1}{2}\frac{dp}{da}\;.
\end{equation}
Clearly, since the energy conservation equation has changed, we cannot use again the GCG equation of state and hope to obtain again the same result as in Eq.~\eqref{rhoc}. What we can do is to take advantage of the foregoing analysis and \textit{assume} a form for $H$ given by Eq.~\eqref{rhoc}. Then, we can derive the equation of state and speed of sound of the corresponding fluid in Rastall's theory and investigate its stability properties.

\begin{figure}[htbp]
 \includegraphics[width=0.4\columnwidth]{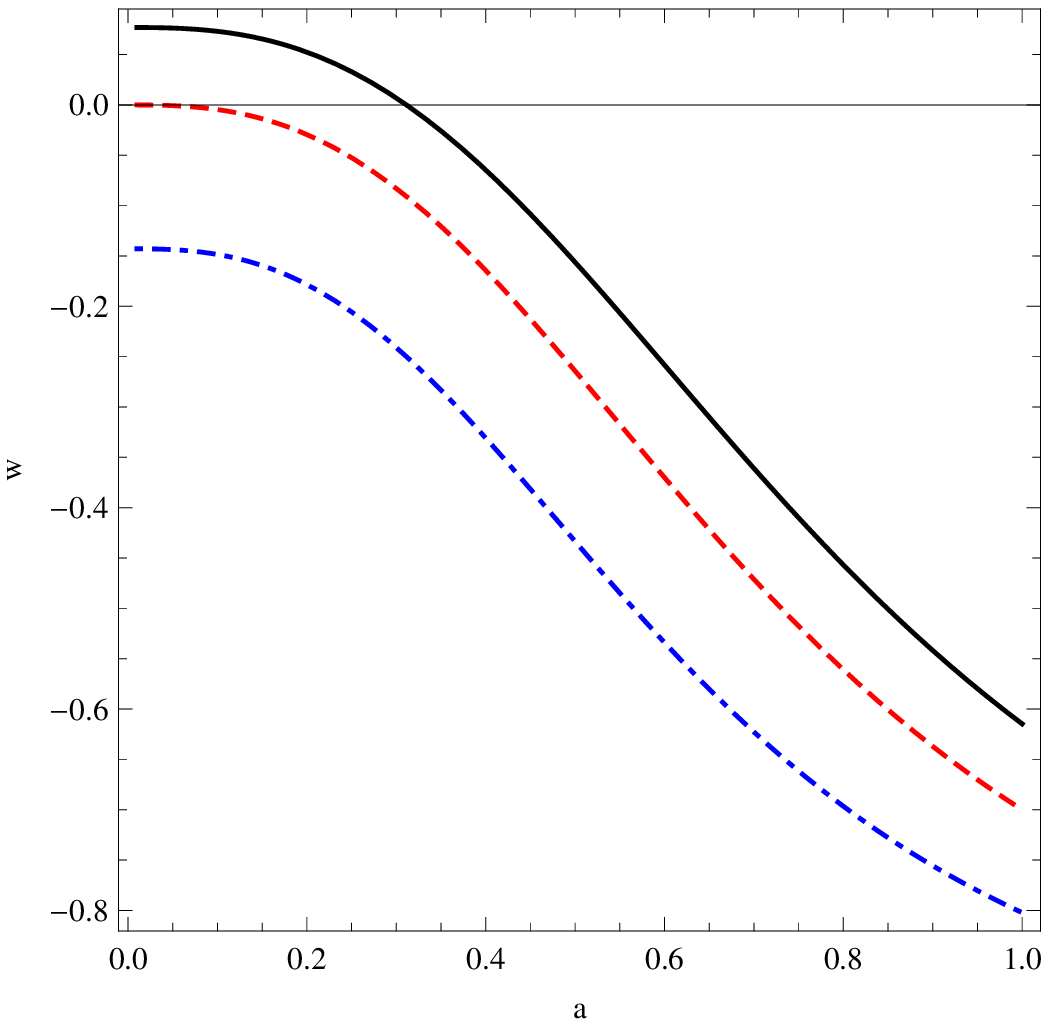}\includegraphics[width=0.4\columnwidth]{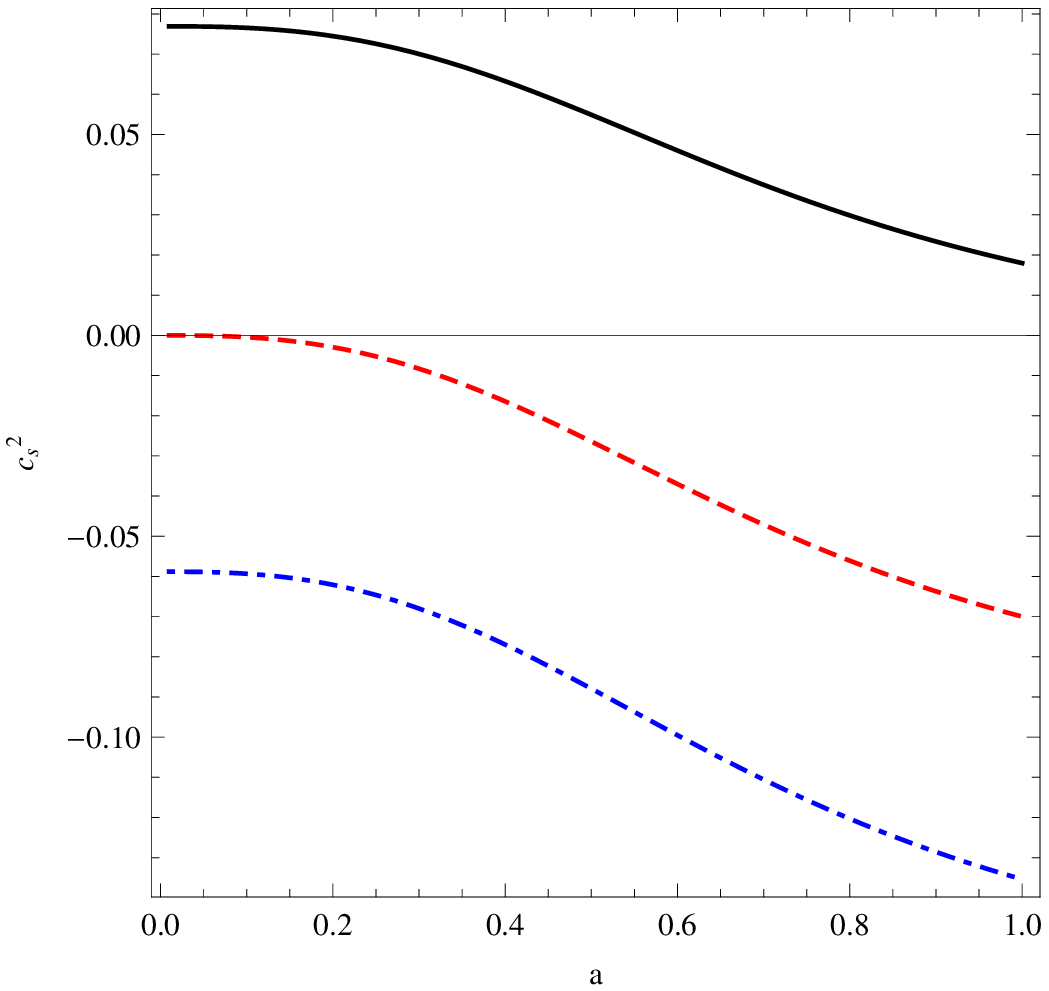}\\
 \caption{Evolution of the equation of state parameter $w$ and of the speed of sound $c_s^2$ given in Eqs.~\eqref{wrastgcg} and \eqref{cs2rastgcg} as functions of the scale factor. The parameter $\gamma$ has been chosen $\gamma = 0.8$ (black solid lines), $\gamma = 1$ (i.e. the GR limit, red dashed lines) and $\gamma = 1.1$ (blue dash-dotted line).}
 \label{Figswcs2}
\end{figure}

Using $E \equiv H^2/H_0^2$, it is easy, by simple inspection of Eqs.~\eqref{friedeqrast} and \eqref{enconsrast} to find
\begin{equation}
 -\frac{3}{a}\left(\rho + p\right) = \frac{dE}{da}\;,
\end{equation}
which is a relation which holds true also for GR. Solving Eq.~\eqref{friedeqrast} for $p$ and substituting in the above equation, one finds
\begin{equation}
 \rho = \frac{\gamma-1}{2(3 - 2\gamma)}a\frac{dE}{da} + \frac{1}{3 - 2\gamma}E\;,
\end{equation}
and for the pressure
\begin{equation}
 p = \frac{\gamma - 3}{6(3 - 2\gamma)}a\frac{dE}{da} - \frac{1}{3 - 2\gamma}E\;.
\end{equation}
Now, as said, we assume
\begin{equation}
 E = \left[\bar A + \frac{(1 - \bar A)}{a^{3(1 + \alpha)}}\right]^\frac{1}{1 + \alpha}\;,
\end{equation}
and then it is straightforward to compute density and pressure. In particular, the equation of state parameter and the speed of sound take the form
\begin{eqnarray}
\label{wrastgcg} w &=& -\frac{2 \bar{A} a^{3 \alpha +3} + (\gamma -1)(1-\bar{A})}{\bar{A}\left(2 a^{3 \alpha +3}+3 \gamma -5\right)-3 \gamma +5}\;,\\
\label{cs2rastgcg} c_s^2 &=& \frac{\bar{A} \left\{\gamma  \left[(\alpha +1) a^{3 \alpha +3}-1\right]-(3 \alpha +1) a^{3 \alpha +3}+1\right\}+\gamma -1}{\bar{A}\left\{3 \gamma  \left[(\alpha +1) a^{3 \alpha +3}-1\right]-(3 \alpha +5) a^{3 \alpha +3}+5\right\}+3 \gamma -5}\;,
\end{eqnarray}
where we have defined the speed of sound as $c_s^2 =dp/d\rho = (dp/da)/(d\rho/da)$, i.e. as if it were adiabatic. We can formally invert the function $\rho(a)$ and then obtain a $p = p(\rho)$, i.e. a barotropic equation of state. We plot in \figurename{~\ref{Figswcs2}} the above $w$ and $c_s^2$ for the choice $\alpha = -0.1$ and $\bar A = 0.7$ (the best fit values found in this paper) and for $\gamma$ varying about unity.

In the range chosen for $\gamma$, the speed of sound is monotonically decreasing and, by inspection, positive if $\gamma \lesssim 0.8$. Let us investigate briefly the asymptotic: in the far past, for $\alpha \to 0$, we have
\begin{eqnarray}
 w &\to& \frac{\gamma - 1}{3\gamma - 5}\;, \qquad a \to 0\;,\\
 c_s^2 &\to& \frac{\gamma - 1}{3\gamma - 5}\;, \qquad a \to 0\;,
\end{eqnarray}
and for $\gamma = 1$ we reproduce the known GR result for the GCG. Now, in the remote future, for $a \to \infty$:
\begin{eqnarray}
 w &\to& -1\;, \qquad a \to \infty\;,\\
\label{cs2rastgcgasympt} c_s^2 &\to& \frac{\gamma - 1 + \alpha(\gamma - 3)}{3\gamma - 5 + 3\alpha(\gamma -1)}\;, \qquad a \to \infty\;,
\end{eqnarray}
where again we reproduce correctly the limit $c_s^2 \to \alpha$ when $\gamma = 1$. By construction, the final stage of the evolution is a de Sitter one. Of course, in all the above calculations we have assumed that $\alpha \ge -1$. Asking for the asymptotic speed of sound of Eq.~\eqref{cs2rastgcgasympt} to be positive, we find the following ranges for $\gamma$:
\begin{equation}
 \gamma < \frac{3\alpha + 1}{1 + \alpha}\;, \qquad \gamma > \frac{5 - 3\alpha}{3(1 - \alpha)}\;.
\end{equation}
For $\alpha = -1$, this amounts to $\gamma < 7/9$ and $\gamma > 53/33$.

\subsection{Metric perturbations and evolution of small perturbations}

We now check the evolution of perturbations.
Let us start by considering the energy-momentum tensor
\begin{eqnarray}
T^{\mu}_{\,\,\nu} &=& \rho u^{\mu}u_{\nu} + p h^{\mu}_{\nu}\;,
\label{EMT}
\end{eqnarray}
where $h^{\mu\,\nu}=g^{\mu\,\nu}+u^{\mu}u^{\nu}$. More explicitly, the background components of (\ref{EMT}) are
\begin{eqnarray}
T^{0}_{\,\,0}=-\rho\;, \quad T^{0}_{i}=T^{i}_{0}=0\;, \quad
T^{i}_{j} = p \delta^{i}_{j} = -\frac{A}{\rho^{\alpha}}\delta^{i}_{j}\, .
\end{eqnarray}
 We assume a conformal Newtonian gauge line element for scalar perturbations
\begin{eqnarray}
ds^{2}=a^{2}\left(\eta\right)\left[-\left(1+2\Phi\right)d\eta^{2}+\left(1-2\Psi\right)\delta_{ij}dx^{i}dx^{j}\right]\;,
\end{eqnarray}
where we introduced the conformal time $\eta$. The perturbations of the fluid 4-velocity up to first order are given by
\begin{eqnarray}
 u^{0}=\frac{1}{a}(1-\Phi), \quad u_{0}=-a(1+\Phi)\;.
\end{eqnarray}
In the absence of anisotropic stresses the spatial off-diagonal Einstein equation implies $\Phi = \Psi$. Assuming the above perturbed metric, we can find the Einstein equation for the potential $\Phi$ for the Rastall's theory, namely
\begin{eqnarray}\label{E00}
 \Delta\Phi - 3\frac{a'}{a}\left(\frac{a'}{a}\Phi + \Phi'\right) = 4\pi G a^2\left[\delta\rho - \frac{\gamma - 1}{2}\left(\delta\rho - 3\delta p\right)\right]\;,\\
\label{Eij} \Phi'' + 3\frac{a'}{a}\Phi' + \left[2\left(\frac{a'}{a}\right)' + \left(\frac{a'}{a}\right)^2\right]\Phi = 4\pi G a^2\left[\delta p + \frac{\gamma - 1}{2}\left(\delta\rho + \delta\rho - 3\delta p\right)\right]\;,
\end{eqnarray}
where the prime denotes derivation wrt the conformal time. Introducing the speed of sound (assuming adiabaticity, i.e. $\delta p/\delta\rho = dp/d\rho$):
\begin{eqnarray}\label{E00bis}
 \Delta\Phi - 3\frac{a'}{a}\left(\frac{a'}{a}\Phi + \Phi'\right) = 4\pi G a^2\delta\rho\left(\frac{3-\gamma}{2} + 3\frac{\gamma - 1}{2}c_s^2\right)\;,\\
\label{Eijbis} \Phi'' + 3\frac{a'}{a}\Phi' + \left[2\left(\frac{a'}{a}\right)' + \left(\frac{a'}{a}\right)^2\right]\Phi = 4\pi G a^2\delta\rho\left(\frac{\gamma - 1}{2} + \frac{5 - 3\gamma}{2}c_s^2\right)\;.
\end{eqnarray}
Changing the conformal time for the scale factor and combining the two above equations, we find an equation for the gravitational potential:
\begin{equation}\label{Phieq}
 \Phi_{aa} + \left(\frac{\mathcal{H}_a}{\mathcal{H}} + \frac{4}{a}\right)\Phi_a + \left(2\frac{\mathcal{H}_a}{\mathcal{H}a} + \frac{1}{a^2}\right)\Phi =
\frac{\gamma - 1 + (5 - 3\gamma)c_s^2}{3 - \gamma + 3(\gamma - 1)c_s^2}\frac{1}{a^2}\left(-\frac{k^2}{\mathcal{H}^2}\Phi - 3\Phi - 3a\Phi_a\right)\;,
\end{equation}
where the subscript $a$ denotes derivation wrt the scale factor, $\mathcal{H} \equiv a'/a$ and we have introduced a plane-wave expansion. Note the ``Rastall factor'' of Eq.~\eqref{Phieq}:
\begin{equation}
 Rf \equiv \frac{\gamma - 1 + (5 - 3\gamma)c_s^2}{3 - \gamma + 3(\gamma - 1)c_s^2}\;.
\end{equation}
Since it multiplies the wavenumber $k$, it may be considered as an ``effective'' speed of sound, different from the adiabatic $c_s^2$ we introduced. In this sense, Rastall's theory seems to call into play a sort of ``geometric entropy''. Using Eq.~\eqref{cs2rastgcg} in order to reduce the ``Rastall factor'' one finds
\begin{equation}
 Rf = \frac{\bar{A}\alpha}{\bar{A} + (1 - \bar{A})a^{-3(\alpha + 1)}}\;,
\end{equation}
which is exactly the GCG speed of sound in GR! Rastall's theory seems to be able to reproduce the same evolution of perturbations of a fluid in GR which provides a given background expansion. Therefore, we may conclude that even framing the GCG in Rastall's theory does not save the model from being ruled out due to the behaviour of small perturbations.

\begin{figure}[p]
 \includegraphics[width=0.4\columnwidth]{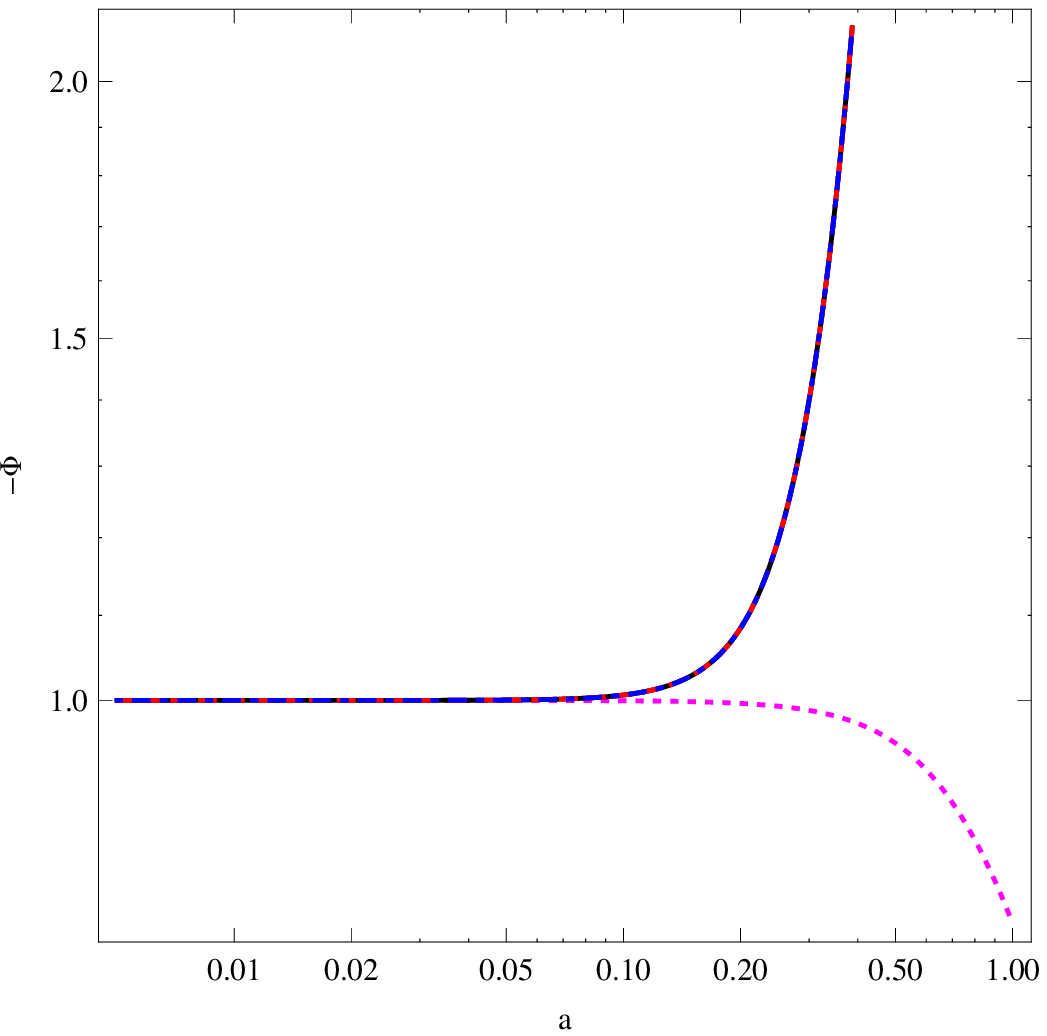}\includegraphics[width=0.4\columnwidth]{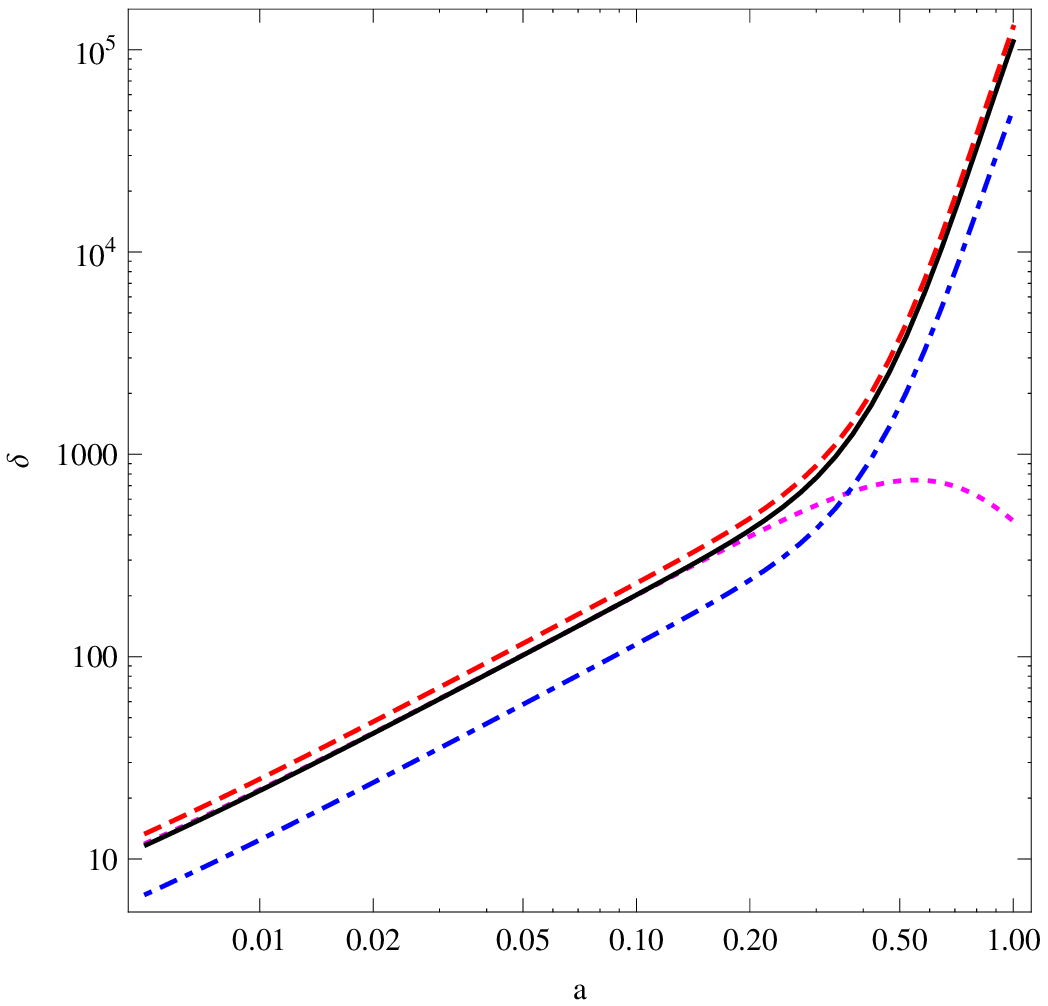}\\
 \caption{Evolution of $\Phi$ and $\delta$ as functions of the scale factor. The GCG parameters have been fixed as $\alpha=-0.1$ and $\bar{A}=0.7$. The parameter $\gamma$ has been chosen $\gamma = 0.5$ (black solid lines), $\gamma = 1$ (i.e. the GR limit, red dashed lines) and $\gamma = 2$ (blue dash-dotted line). For comparison, the $\Lambda$CDM lines are depicted as the magenta dotted ones.}
 \label{FigsPhidelta}
\end{figure}

Just to check. We choose as initial conditions $\Phi = -1$ and $\Phi_a = 0$ in $a = 0.005$ and let the system evolve with the choice $\alpha = -0.1$ and $\bar{A} = 0.7$. In \figurename{~\ref{FigsPhidelta}} we plot the evolution of $\Phi$ and $\delta$ as function of the scale factor for a representative scale $k = 0.1$ h Mpc$^{-1}$.

Clearly, the results are catastrophic. A remaining possibility of salvation is to adopt a scalar field description for the GCG, as done in \cite{thais}. As shown in \cite{Gao:2009me} (where the authors do not mention Rastall's theory and investigate just the $\gamma = 2$ case) and \cite{Daouda:2012ig} in a scalar field approach the effective speed of sound in the rest frame of the field (i.e. where $T_0^i = 0$) can be written as
\begin{equation}
 c_s^2 = \frac{2 - \gamma}{\gamma}\;,
\end{equation}
which evidently vanishes for $\gamma = 2$. Since the scalar field description possesses one degree of freedom more than the fluid description (encoded in the scalar field potential) the background expansion can be fixed without any drawbacks on the evolution of perturbations, which for $\gamma = 2$ may be identical to the one of a pressureless fluid in GR (e.g. CDM).

\section{Evolution of sub-horizon perturbations}\label{Sec:meszeq}

The results of our background tests do not impose negative values for $\alpha$ with a high confidence level. This could mean that the GCG has indeed $\alpha=0$ meaning that the dark sector of the universe is actually an ordinary adiabatic fluid with a small, constant and negative pressure. Hence, we can argue that at high redshifts the GCG does not look exactly like standard CDM ($p_{cdm}=0$).

Our goal in this section is to revisit the issue concerning the equivalence of GCG with $\alpha=0$ and the $\Lambda$CDM model. Obviously both cases produces the same background expansion. However, does this equivalence holds at first order in the perturbations?

We present an analysis of the M\'esz\'aros effect which describes the formation of dark matter structures during the first stages of the matter dominated epoch. We assume a single fluid description, i.e. we neglect the contribution of baryons and radiation. This is a reasonable approximation if we want to track the growth of sub-horizon perturbations during the matter dominated epoch. For the technical details, we follow the analysis of \cite{dominik}.

For the linear perturbations of (\ref{EMT}) we define the velocity scalar $v$, which is associated with the peculiar velocity by $\delta u^{i}{}_{,i} \equiv kv/a$.  At first order, the (0-0) component of the Einstein equation reads
\begin{eqnarray}
-k^{2}\Psi-3\mathcal{H}\Psi^{\prime}-3\mathcal{H}^{2}\Psi = \frac{3}{2} \mathcal{H}^2
\delta_{\rm tot}\;,
\label{poisson}
\end{eqnarray}
where $\delta_{\rm tot}$ is the \textit{total} density contrast, i.e. $\delta_{\rm tot} \equiv \delta\rho_{\rm tot}/\rho_{\rm tot}$. During the time interval between the kinetic decoupling of DM particles from the primordial plasma and the epoch of matter radiation equality ($z_{eq}\approx 3300$ for the $\Lambda$CDM model) the sub-horizon DM perturbations grow only logarithmically with the scale factor. After $z_{eq}$ the DM perturbations obey to $\delta\propto a$. This is the main result behind the so called M\'esz\'aros effect \cite{Meszaros:1974tb}.

We study here an example in which the GCG with $\alpha = 0$ behaves differently from the $\Lambda$CDM. We show that the growth of sub-horizon GCG matter perturbations is different from the standard CDM. In order to obtain a M\'esz\'aros-like equation for the GCG we make use of the covariant conservation of the energy-momentum tensor ($T^{\mu}{}_{\nu;\,\mu}=0$). At first-order we find
\begin{eqnarray}
\delta^{\prime} = -3\mathcal{H}\delta\left(c^{2}_{s} - w_c\right) -
\left(1+w_c\right)\left(k v - 3\Psi^{\prime}\right)\;,
\label{cons1}
\end{eqnarray}
and
\begin{eqnarray}
v^{\prime} = -\mathcal{H}\left(1 - 3w_c\right) v - \frac{w_c^\prime}{1 + w_c}v
+ k \Psi + \frac{k c^{2}_{s}}{1 + w_c} \delta\;,
\label{cons2}
\end{eqnarray}
for the energy and momentum balances of each single component (which we assume conserving separately), respectively.

For small scales we can neglect the term $\Psi^{\prime}$ in Eq. (\ref{cons1}) and we also take the sub-horizon limit of the Poisson equation Eq. (\ref{poisson}). Together with Eq. (\ref{cons2}) and with the fact that $\delta_{\rm tot} = \delta_{gcg}$, we find a M\'esz\'aros-like equation for the GCG with $\alpha=0$, i.e. $c^2_s=0$:
\begin{eqnarray}
a^{2}\frac{\,d^{2}\delta_{gcg}}{da^{2}}+\left[3(1-w_c)+\frac{a}{H}\frac{\,d
\,H}{da}\right]a\frac{\,d\delta_{gcg}}{da}+ \nonumber\\
\left[-\frac{3}{2}-\frac{15}{2}w_c+9w_c^2-\frac{3aw_c}{H}\frac{dH}{da}-3a\frac{dw_c}{da}\right]\delta_{gcg}=0\;.
\label{small}
\end{eqnarray}
The standard equation for CDM in the $\Lambda$CDM model can be obtained in the same way, but remembering that now 
\begin{equation}
 \delta_{\rm tot} = \frac{\rho_{cdm}}{\rho_{cdm} + \rho_\Lambda}\delta_{cdm} = \Omega_{cdm}\delta_{cdm}\;,
\end{equation}
since the $\Lambda$CDM model is basically a two-fluid model. Therefore, we have
\begin{equation}
a^{2}\frac{\,d^{2}\delta_{cdm}}{da^{2}}+\left[3+\frac{a}{H}\frac{d
\,H}{da}\right]a\frac{\,d\delta_{cdm}}{da}-\frac{3}{2}\Omega_{cdm}\delta_{cdm}=0\;.
\label{smallcdm}
\end{equation}
Identifying the background expansion of the two instances, $\Lambda$CDM and GCG, one can write $\Omega_{cdm} = 1 + w_c$ and thus
\begin{equation}
a^{2}\frac{\,d^{2}\delta_{cdm}}{da^{2}}+\left[3+\frac{a}{H}\frac{d
\,H}{da}\right]a\frac{\,d\delta_{cdm}}{da}-\frac{3}{2}(1 + w_c)\delta_{cdm}=0\;.
\label{smallcdm2}
\end{equation}
Equations \eqref{small} and \eqref{smallcdm2} are clearly different (note that $H$ and $w_c$ have the same evolution), therefore $\delta_{gcg}$ and $\delta_{cdm}$ evolve differently. Indeed, one can show that $\delta_{gcg} = (1 + w_c)\delta_{cdm}$. This is also a result of \cite{martins}, see also \cite{Avelino:2007tu, Avelino:2008cu, Aviles:2011ak}.

\begin{figure}[h]
 \centering
\includegraphics[width=0.4\columnwidth]{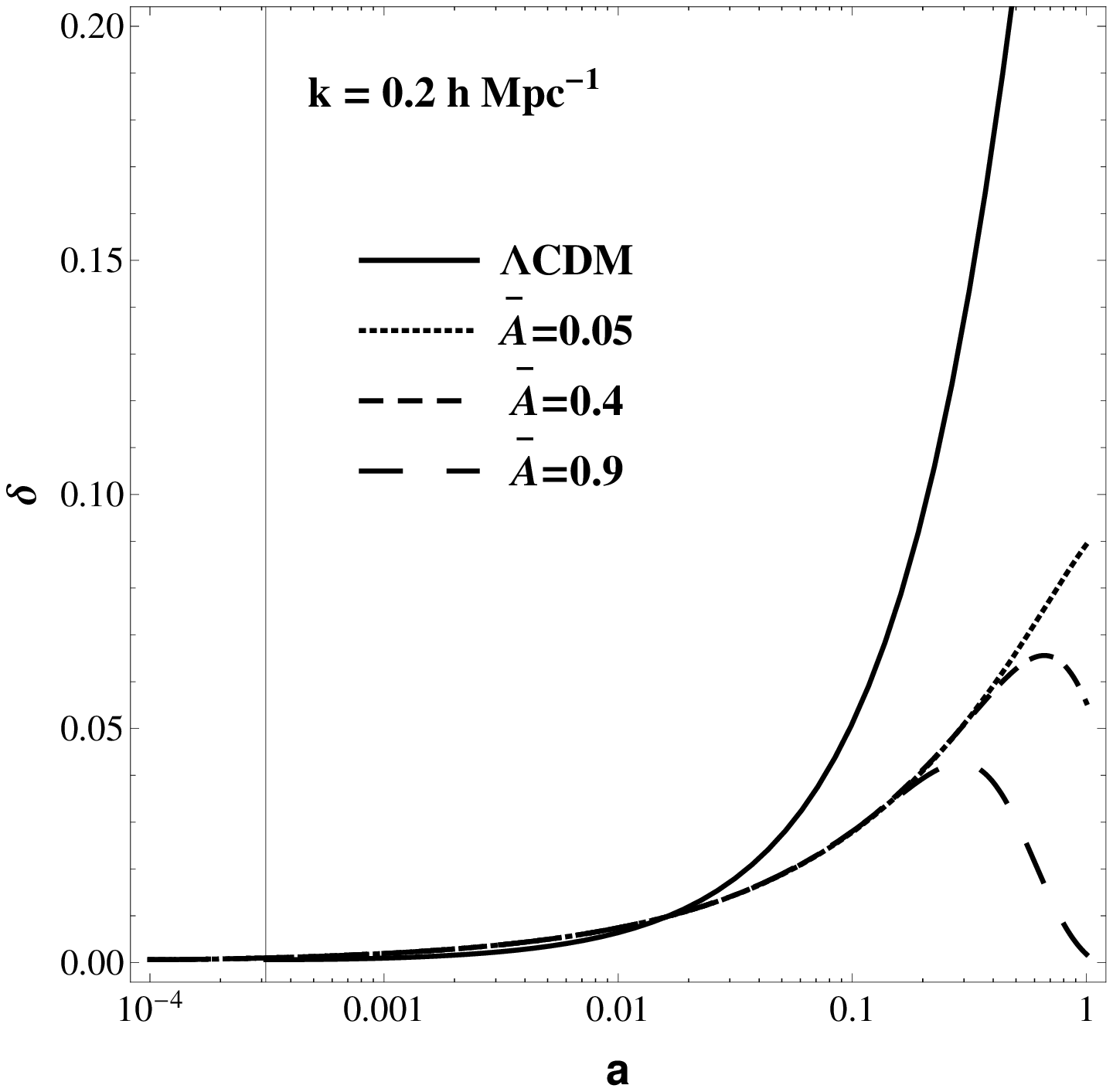}\includegraphics[width=0.4\columnwidth]{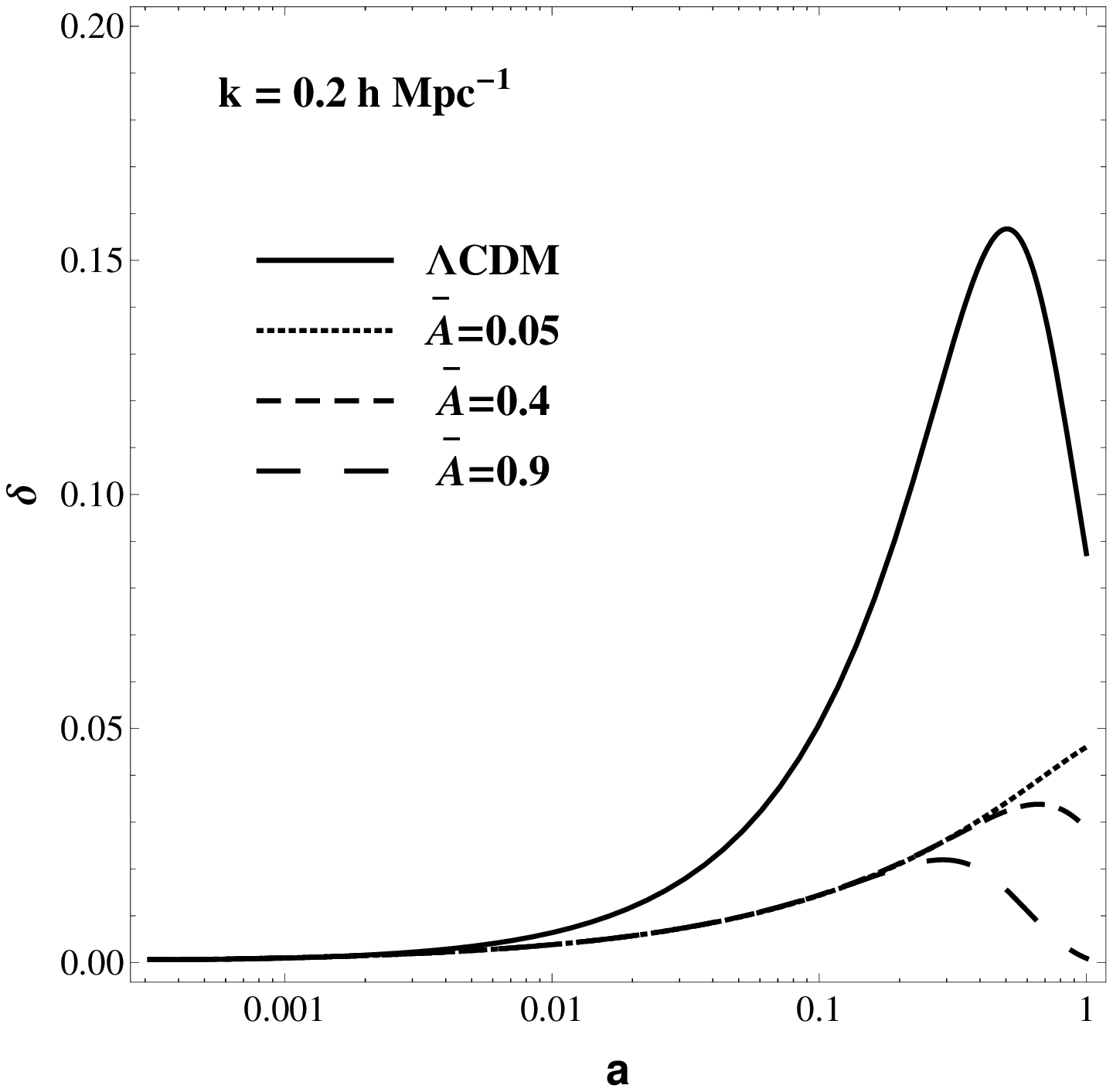}
 \caption{Evolution of the sub-horizon density contrast $\delta$. The standard pressureless CDM ($p = 0$) is shown as the solid line. For the GCG with $\alpha=0$ we show the evolution of $\delta$ for three different values of $\bar{A}$. Left panel: $\delta_{cdm} \equiv \delta\rho_{cdm}/\rho_{cdm}$. Right panel: $\delta_{cdm} \equiv \delta\rho_{cdm}/(\rho_{cdm} + \rho_\Lambda)$.}
 \label{Fig9}
\end{figure}

We show in \figurename{~\ref{Fig9}} the evolution of a typical sub-horizon scale after the equality ($z_{eq} \approx 3000$). We assume that both CDM and GCG have the same initial conditions at that time. We set them by generating the power spectrum at that time for the scale $k = 0.2$ h Mpc$^{-1}$ with help of the CAMB code \cite{Lewis:1999bs}. We then solve numerically Eq. \ref{small} for $\bar{A}=0.05,0.4$ and $0.90$. The perturbations in the GCG fluid are strongly suppressed in comparison to the CDM case. This result can also be appreciated in Fig.~1 of \cite{Aviles:2011ak}.

The main conclusion of \cite{martins} is that the $\Lambda$CDM model and the $\alpha = 0$ GCG are indistinguishable. We agree on this at the linear perturbations level since, being the speed of sound vanishing, the evolution of the gravitational potential $\Psi$ is determined univocally and it is the same in the two models (it depends on the background evolution only, which is identical in the two models, by construction). This imply that the evolution of $\delta_{\rm tot}$ is also the same and therefore, if the components conserve independently, the evolution of the density contrasts of baryons and radiation are also respectively identical in the two models. Since the latter two are the only observable components, one may conclude that the $\Lambda$CDM model and the $\alpha = 0$ GCG are indistinguishable, from gravity only. No difference can be expected for the e.g. integrated Sachs-Wolfe effect or for the baryonic large scale structure power spectrum (which is indeed the one we infer from observation). On 
the other hand, the density contrast $\delta_{gcg}$ seems unable to reach the non-linear regime $\delta\approx 1$, see \figurename{~\ref{Fig9}}. We wonder if this would not have a dramatic effect on the halo formation. The absence of the latter would be in contrast with the observation of the almost flat velocity curves of galaxies and with lensing experiments. Probably, only numerical simulations would provide the correct matter distribution, but semi-analytical studies touching the non-linear regime \cite{neven, fernandes} could supply additional information about the final fate of the $\alpha = 0$ GCG.

\section{Conclusions}\label{Sec:Conc}

We performed a Bayesian analysis of the background behaviour of the GCG model using $H(z)$, SNIA, CMB and BAO datasets. We focused particularly on the parameter $\alpha$, on which a huge literature already exists. Our result is that $\alpha = - 0.087^{+0.159}_{-0.135}$, at the 2$\sigma$ level, i.e. negative values of $\alpha$ seem to be favoured over the positive ones. Indeed $\alpha$ is negative with 85\% of confidence. The uncertainty is still too large for us to claim that a ``tension'' with perturbative tests (which constrain $|\alpha| \lesssim 10^{-4}$ at 2$\sigma$) does exist, but we speculate on the consequences of this occurrence, which is not far from being settled, given the ongoing and forthcoming observational programs which collect more precise data day after day. That kind of inconsistency would immediately rule out the GCG model, since it is unacceptable to obtain two different set of parameters depending on the observational test applied. We assume this to occur and figure out how to possibly 
save the GCG unification paradigm. We introduce a fluid
model in Rastall's theory which exactly behaves as the GCG at the background level (i.e. it provides the same evolution for $H(z)$) and discuss its perturbative properties. It turns out that small and positive speeds of sound are compatible with $\alpha < 0$. On the other hand, we show that the evolution of perturbations is ruled by an effective speed of sound which is identical to the GCG one in GR, as if Rastall's theory introduced a sort of ``geometric entropy''. Therefore, as one should expect, the results are catastrophic, with the density contrast and the gravitational potential growing too fast (for $\alpha < 0$) for being in agreement with observation. As a possible salvation, we indicate a scalar field description of the GCG in Rastall's theory, where the background expansion and the effective speed of sound can be fixed independently (and the latter to zero).

We also address the issue of the $\alpha \to 0$ limit. It is indistinguishable from the $\Lambda$CDM model at the linear perturbative regime, but we show that the small negative pressure of the GCG affects the evolution of the GCG density contrast during the matter dominated phase, where it is expected to behave as CDM and provides a different evolution for the sub-horizon matter perturbations. This analysis shows that sub-horizon structures (e.g. dark halos) may not form as in the standard $\Lambda$CDM case. Thus, it could be very interesting to perform numerical simulations for matter fluids with a hydrodynamical evolution distinct from the standard pressureless case. This analysis could clarify if the $\alpha \to 0$ limit of the GCG is really identical to the $\Lambda$CDM model.


\begin{acknowledgements}
This work was supported by CNPq (Brazil). HV acknowledges support from the DFG within the Research Training Group 1620 ``Models of Gravity''. We would like to thank Alejandro Aviles and Pedro Avelino for enlightening remarks and suggestions.
\end{acknowledgements}

\end{document}